\documentclass[journal]{IEEEtran}
\usepackage[linesnumbered,lined,ruled,commentsnumbered]{algorithm2e}
\usepackage{threeparttable}
\usepackage{tcolorbox}
\usepackage{amsmath}
\usepackage{orcidlink}
\usepackage{paralist}
\usepackage{makecell}
\usepackage{colortbl}
\usepackage{xcolor}
\usepackage{xspace}
\usepackage{multirow}
\usepackage{array}
\usepackage{subfigure}
\usepackage{pifont}
\usepackage{algorithm2e}
\usepackage{multicol}
\usepackage{algorithmic}
\usepackage[numbers]{natbib}
\usepackage{orcidlink}
\usepackage{url}
\usepackage{cleveref}
\usepackage{booktabs}

\newcommand{\APIGH}{\textsc{APISensor}\xspace}

\hypersetup{hidelinks}

\newtcolorbox{summarybox}[1][summaries]{
  colback=white,
  colframe=gray!50!black,
  title=#1,
  attach a boxed title to bottom center,
  boxed title style={augmented, boxsep=0.1mm, top=0.1mm, bottom=0.1mm},
  overlay={
    \draw[gray] (title.south) -- (frame.south);
  }
}

\DeclareFontFamily{OT1}{pzc}{}
\DeclareFontShape{OT1}{pzc}{m}{it}{<-> s * [1.10] pzcmi7t}{}
\DeclareMathAlphabet{\mathpzc}{OT1}{pzc}{m}{it}

\definecolor{darkgrey}{HTML}{434343}
\definecolor{lightgrey}{HTML}{A9A9A9}
\definecolor{silver}{HTML}{C0C0C0}
\definecolor{midgrey}{HTML}{808080}
\definecolor{white}{HTML}{FFFFFF}
\colorlet{gray1}{gray!70}
\colorlet{gray2}{gray!30}
\colorlet{gray3}{gray!15}
\colorlet{gray4}{gray!50}
\colorlet{gray5}{gray!20}

\AtBeginDocument{%
  \providecommand\BibTeX{{%
    \normalfont B\kern-0.5em{\scshape i\kern-0.25em b}\kern-0.8em\TeX}}}
    
\begin{document}

\title{\APIGH: Robust Discovery of Web API from Runtime Traffic Logs} 



\author{Yanjing Yang\IEEEauthorrefmark{2}$^{\orcidlink{0009-0006-8789-4589}}$,
Chenxing Zhong\IEEEauthorrefmark{2}$^{\orcidlink{0000-0001-7298-1926}}$,
Ke Han\IEEEauthorrefmark{2}$^{\orcidlink{0009-0001-5094-4427}}$,
        Zeru Cheng$^{\orcidlink{0009-0001-5094-4427}}$,
        Jinwei Xu$^{\orcidlink{0009-0004-5157-1118}}$,
        Xin Zhou\IEEEauthorrefmark{1}$^{\orcidlink{0000-0002-3263-1275}}$,

        He Zhang\IEEEauthorrefmark{1}$^{\orcidlink{0000-0002-9159-5331}}$,
        Bohan Liu$^{\orcidlink{0000-0002-0146-5411}}$,
        
\thanks{
\IEEEcompsocthanksitem 
\IEEEauthorrefmark{2} Contribute Equally
\IEEEcompsocthanksitem 
\IEEEauthorrefmark{1} 
\textbf{Corresponding author: Xin Zhou \&  He Zhang.}
\IEEEcompsocthanksitem Yanjing Yang, Ke Han, Xin Zhou, Jinwei Xu, Bohan Liu, and He Zhang are with State Key Laboratory of Novel Software Technology, Software Institute, Nanjing University, Nanjing 210008, China 
(e-mail: yj\_yang@smail.nju.edu.cn; zhouxin@nju.edu.cn; bohanliu@nju.edu.cn; hezhang@nju.edu.cn; jinwei\_xu@smail.nju.edu.cn;.
\IEEEcompsocthanksitem Chenxing Zhong is with Nanjing University of Science and Technology, Nanjing 2100094, China
    }}

\maketitle

\begin{abstract}

Large Language Model (LLM)–based agents increasingly rely on APIs to operate complex web applications, but rapid evolution often leads to incomplete, unclear, or inconsistent API documentation. 
Existing work broadly falls into two categories: (1) static, white-box approaches based on source code or formal specifications, and (2) dynamic, black-box approaches that infer APIs from runtime network traffic. Static approaches rely on internal artifacts such as source code or API specifications, which are typically unavailable for closed-source systems. Moreover, because they analyze potential interfaces rather than runtime behavior, they often over-approximate API usage by matching patterns that may include unused or unreachable code, resulting in high false-positive rates.
Although dynamic black-box API discovery applies broadly across applications, its robustness degrades in complex traffic environments where shared collection points aggregate traffic from multiple applications.
To improve the robustness of API discovery under mixed runtime traffic, we propose \APIGH, a black-box API discovery framework that reconstructs application APIs in an unsupervised setting.
Instead of relying on simple clustering or analyzing requests individually, \APIGH performs structured analysis over complex traffic, combining traffic denoising and normalization with a graph-based two-stage clustering process to recover accurate APIs.
Our evaluation measures the accuracy of \APIGH in discovering APIs and its robustness across six widely used web applications using over 10,000 runtime requests with simulated mixed-traffic noise. The results demonstrate that \APIGH significantly improves discovery accuracy, achieving an average Group Accuracy Precision of 95.92\% and an F1-score of 94.91\%, outperforming state-of-the-art API discovery methods. Across different applications and noise settings, \APIGH achieves both the lowest performance variance and at most an 8.11-point FGA drop, demonstrating the best robustness in 10 baselines. Finally, ablation studies confirm that each component of the \APIGH is essential to its overall effectiveness. During our experiments, \APIGH also reveals API documentation inconsistencies in a real application, which are confirmed by the developers in the community.



\end{abstract}


\section{Introduction}

The rapid growth of Large Language Models (LLM) based agent applications (e.g., Openclaw~\cite{basu2026openclaw}) has increased the requirements for reliable API discovery. Agents allow LLM to operate complex web applications by invoking APIs without requiring knowledge of UI design logic~\cite{wolflein2025agenttools, wang2025codesync}. However, many web applications still provide incomplete or missing API descriptions because API implementations evolve rapidly through frequent version updates and feature iterations, while documentation maintenance often lags behind the implementation~\cite{lin2023detecting, shi2011empirical, zhong2013detecting}.

Existing API discovery techniques can be broadly divided into (1) white-box and (2) black-box approaches, which can work in synergy in practice. White-box approaches that rely on source code or static specifications have limited practical applicability because they require access to program internals. However, most production web services are proprietary and closed source. Recent industry reports estimate that more than 85\% of deployed web services are not publicly available in source form, which significantly restricts the use of white-box techniques \cite{huang2024generating}. Moreover, many white-box approaches rely on static analysis, which typically adopts \textit{over-approximation} to conservatively model program behavior. As a result, the analysis may report APIs that are defined in the code but not actually exposed at runtime (e.g., endpoints used only for internal service calls), leading to false positives in discovered active API assets \cite{huang2024generating}. In addition, prior work shows that existing documentation may not accurately reflect the actual implementation \cite{wang2024dainfer}. Therefore, understanding application functionality cannot rely solely on white-box techniques and requires black-box runtime behavior observation~\cite{luo2024holistic}.


Although black-box API discovery techniques are applicable to a wider range of scenarios, their effectiveness often degrades significantly in practice when operating on diverse runtime traffic. Existing approaches mainly follow three directions, including traffic capture methods that (1) reconstruct APIs from request–response traces~\cite{mitmproxy2swagger, optic}, (2) request pattern analysis using clustering or learning-based techniques~\cite{liu2020webapisearch, yandrapally2023apicarv, Gu2016deep}, and (3) log template mining approaches that extract parameterized request structures~\cite{nedelkoski2020log, wang2019attributed, vaarandi2015logcluster, dai2022logram, liu2022uniparser, logppt}. In practice, API traffic is commonly collected at shared network gateways that serve multiple applications \cite{han2024byways}. As a result, requests from different services are interleaved with each other and with background traffic \cite{wallner2024feature}. For example, a gateway may simultaneously observe requests such as \texttt{/account/info/get}, \texttt{/accounting/info/gen}, and \texttt{/account/info/graph}. Although these paths share highly similar lexical patterns, they may belong to user account management, financial accounting, and analytics services, respectively. The same gateway may also carry non-API traffic such as \texttt{/static/account.js} or \texttt{/images/account.png}. Such heterogeneous traffic makes it difficult to correctly attribute observed requests to the application that actually exposes the corresponding API \cite{wang2024diagnosing, vanede2020flowprint}.
\textbf{\textit{(Problem)} Consequently, existing black-box traffic-based API discovery approach are not robust and suffer substantial performance degradation in complex traffic environments involving multiple applications.}

\textbf{\emph{(Approach)}}
To address the above challenges, we propose \APIGH, a black-box API discovery framework that extracts structural API templates from runtime traffic and clusters them using graph-based analysis.
To cope with mixed traffic from multiple applications and heavy noise in gateway environments (Challenge~1), \APIGH first filters runtime traffic to remove non-API requests and normalizes request paths to reduce noise and variability. This step suppresses interference from background traffic and produces cleaner inputs for subsequent analysis.
To overcome the limited information available from individual requests (Challenge~2), \APIGH performs API discovery in two stages. Structural template extraction abstracts variable identifiers while preserving stable API structures, which mitigates fragmentation caused by dynamic parameters. Based on these templates, graph-based clustering captures relationships among API calls rather than treating requests independently, enabling more reliable API discovery under unlabeled and heterogeneous traffic.

\textbf{\emph{(Evaluation)}}
To evaluate \APIGH, we conduct API discovery experiments on several well-known web applications, including Train-Ticket, HumHub, Memos, Overleaf, Nextcloud, and Dify. We use \emph{Precision of Group Accuracy} (\textbf{PGA}), \emph{Recall of Group Accuracy} (\textbf{RGA}), the \emph{F1-score of Group Accuracy} (\textbf{FGA}), and Purity as evaluation metrics. Averaged across all projects, \APIGH achieves leading performance with a \textbf{PGA} of 95.92\%, an \textbf{RGA} of 94.36\%, and an \textbf{FGA} of 94.91\%, while maintaining low variance across different applications. These results demonstrate strong robustness across projects and a low false-positive rate in practical API discovery. Then, we further evaluate the robustness of \APIGH under different noise types and injected noise levels. Across all settings, \APIGH shows stronger robustness than all baseline methods and is minimally affected by noise introduced in simulated gateway environments. Across different noise types, the performance variation of \APIGH remains limited, with the \textbf{FGA} decreasing by at most 8.11\% and Purity varying within 1\%. Under different noise levels, the performance variation remains within 4\%. Finally, we conduct ablation studies on the key components of \APIGH, showing that each design component is necessary for achieving the overall performance.

The contributions of this study are summarized as follows.
\begin{itemize}
\item \textbf{Data:} We construct a benchmark dataset by manually collecting over 10K runtime traffic traces from six web applications of different types, scales, and domains. Based on official documentation and validation through real execution traffic, we annotate 199 valid and callable API endpoints for evaluation. The dataset will be released as part of our replication package~\cite{}.

\item \textbf{Approach:} We propose \APIGH, a black-box API discovery framework that reconstructs web APIs from mixed runtime traffic. The framework combines traffic denoising, structural API template extraction, and robust graph clustering to achieve accurate and stable API discovery under unsupervised and noisy settings.

\item \textbf{Evaluation:} We comprehensively evaluate \APIGH against existing baseline methods under the identified challenges, covering diverse applications and different noise conditions. In addition, during evaluation, \APIGH uncovers previously undocumented shadow APIs in a popular application, Dify, which are later confirmed by the developer community.
\end{itemize}

The remainder of this paper is organized as follows.
\Cref{sec:relatedwork} reviews related work and motivates our work by identifying unresolved challenges.
\Cref{sec: approach} presents our \APIGH approach for API discovery from runtime traffic.
\Cref{sec:experiment design} describes the research questions and the experimental setup.
\Cref{sec: Result analysis} reports and analyzes the results.
\Cref{sec: Discussion} discusses the value of our approach for agent-based applications and the shadow undocumented APIs discovered in Dify.
Finally, we examine threats to validity in \Cref{sec:TTV} and conclude this paper in \Cref{sec: conclusion}.

\section{Related Work}
\label{sec:relatedwork}


Prior work on API reverse engineering and discovery mainly falls into two categories:
(1) \textbf{white-box approaches} that analyze \textbf{source code} and other development-time artifacts to statically extract API definitions, which provide broad coverage but are limited to open-source systems; and
(2) \textbf{black-box approaches} that infer APIs from \textbf{runtime traffic} and logs, which reflect actual exposed functionality but often rely on fixed assumptions about API structures.
These two lines of work are complementary, as static analysis improves completeness while traffic-based analysis enhances accuracy, together enabling more comprehensive and precise API discovery.

White-box API discovery approaches analyze artifacts available at development time, such as API specifications, source code, and developer documentation.
Specification-driven methods, for example OpenAPI, describe intended API interfaces in a structured way and support systematic analysis~\cite{openapi_spec}.
However, specifications may become outdated as systems evolve, leading to specification drift and the existence of undocumented or obsolete APIs~\cite{espinha2014web}.
In addition, inconsistent REST design practices reduce the reliability of specification-based analysis~\cite{Murphy2017}. To reduce manual effort, some approaches extract API information directly from source code and repositories.
Tools such as APIDocGen~\cite{Nybom2018} and RepoDocGen~\cite{8530111} recover API endpoints and structures through static code analysis.
These approaches can achieve good coverage when full source code is available, but they only reflect design-time artifacts and do not capture actual runtime API usage~\cite{learning_glossary}.
Other work focuses on improving API discovery by linking developer intent to API usage.
TaskAPIRec~\cite{Huang2018} and related studies~\cite{McMillan2011} map high-level tasks to API calls, while CodeGlossary~\cite{learning_glossary} extracts semantic information from code.
Additional semantic knowledge has also been collected from developer forums~\cite{Chen2014,Rahman2016}.
More recent hybrid approaches, such as gDoc, combine static analysis with limited runtime information to generate structured API documentation~\cite{wang2023gdoc}.

\begin{figure}[t]
    \centering
    \includegraphics[width=1.0\columnwidth]{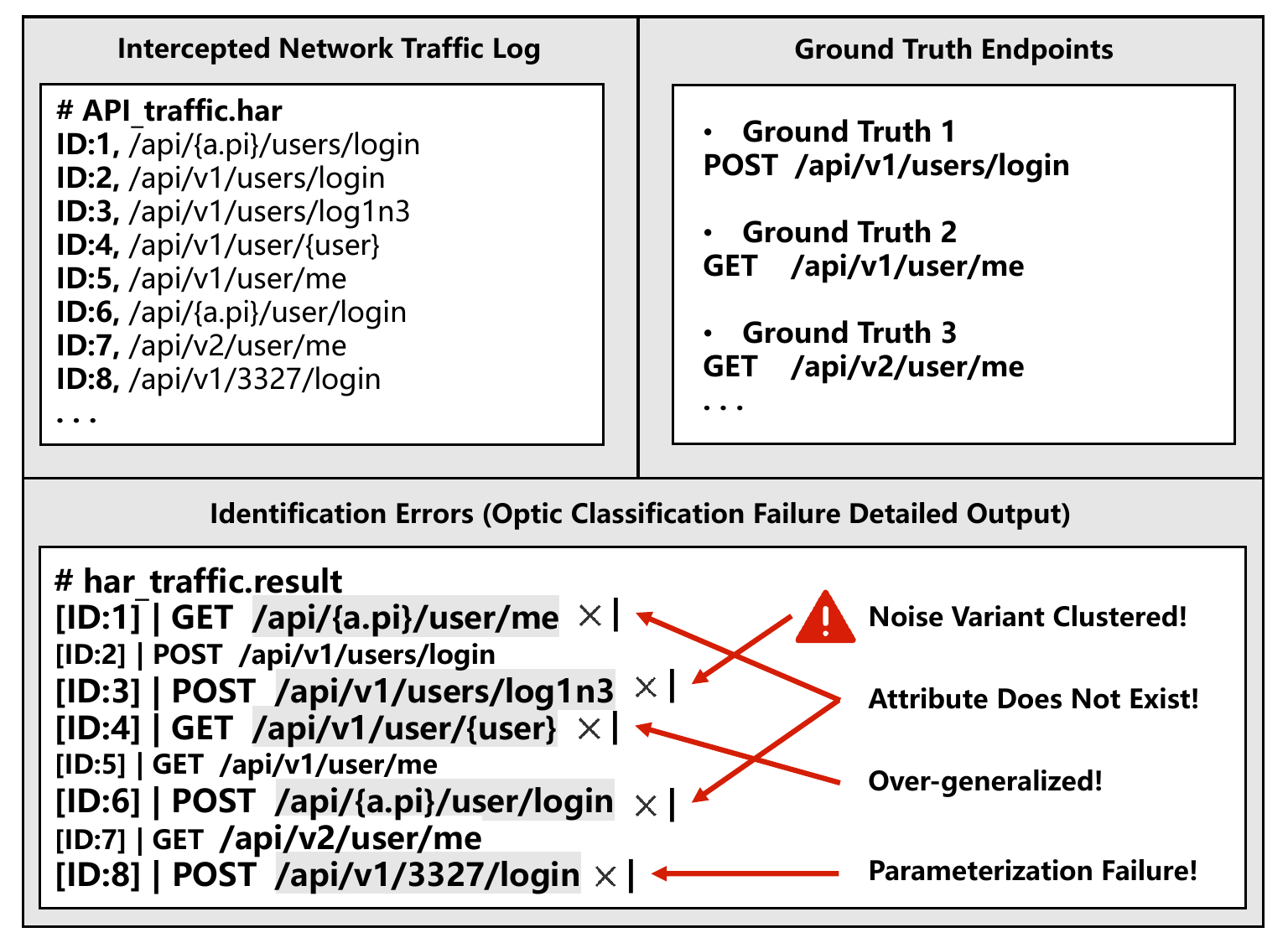}
    \caption{Existing tools (e.g., Optic)  cannot reliably distinguish these heterogeneous requests, potentially leading to incomplete and inaccurate API
specifications.}
    \label{fig:motivation_optic}
\end{figure}

While white-box approaches rely on development-time artifacts, API discovery can also be performed from a runtime perspective. Accordingly, black-box approaches infer API endpoints and behaviors directly from observed traffic and logs, without access to internal artifacts.

Several open-source tools support API reverse engineering from runtime traffic. Optic~\cite{optic} and Mitmproxy2Swagger~\cite{mitmproxy2swagger} generate OpenAPI specifications from HTTP request--response traces or intercepted network traffic, but they mainly rely on heuristics and shallow structure inference. Beyond such tools, prior work explicitly studies endpoint discovery and specification recovery using richer signals. APID2Spec~\cite{yang2018extracting} extracts API specifications by crawling documentation pages and inferring base URLs, paths, and HTTP methods. APICARV~\cite{yandrapally2023apicarv} monitors UI-driven runtime traffic and builds an interaction graph to infer endpoints. Web API Search~\cite{liu2020webapisearch} supports endpoint-level discovery by matching natural language queries against API descriptions mined from online documentation. Online methods such as APIDrain3 incrementally mine URL templates from streaming API traffic for continuous discovery.

Template mining and log parsing techniques are also related and can be applied to API traffic analysis. Classical parsers such as LogCluster~\cite{vaarandi2015logcluster} and LogNgram~\cite{dai2022logram}, as well as more recent approaches like UniParser~\cite{liu2022uniparser} and LogPPT~\cite{logppt}, extract stable patterns by abstracting variable tokens from request paths. However, since these methods mainly rely on lexical patterns of individual requests, they often struggle to distinguish API operations with similar URL structures.

Figure~\ref{fig:motivation_optic} shows a real example of Optic on gateway traffic, highlighting its limitations in accurate API identification. The intercepted traffic contains multiple variants of similar paths (e.g., \texttt{/api/v1/users/login}, \texttt{/api/v1/users/log1n3}, \texttt{/api/\{a.pi\}/user/login}) together with noisy or malformed requests. While the ground truth includes canonical endpoints such as \texttt{POST /api/v1/users/login} and \texttt{GET /api/v1/user/me}, Optic produces erroneous outputs such as invalid attributes (e.g., \texttt{/api/\{a.pi\}/user/me}), over-generalized paths, clustered noise variants, and incorrect parameterization (e.g., \texttt{/api/v1/3327/login}). 

These errors stem from two key challenges: \textbf{(1) Mixed and noisy gateway traffic}, where requests from different API versions and malformed paths coexist in a single stream; and \textbf{(2) Insufficient context in individual requests}, as many requests share similar lexical patterns but differ semantically, leading to over-generalization and misclassification when analyzed in isolation.

\begin{figure*}[t]
\centering
\includegraphics[width=1.0\linewidth]{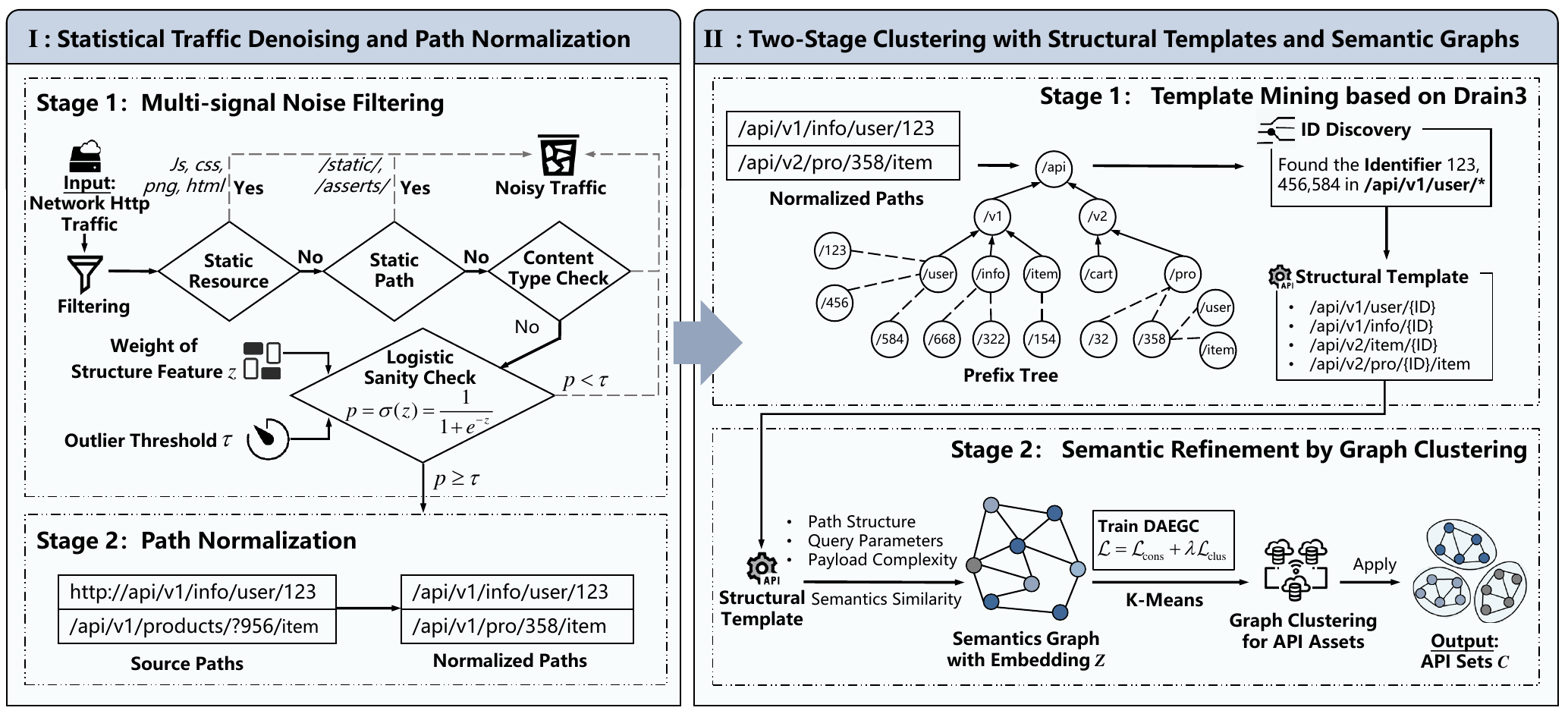}
\caption{The workflow of \APIGH.}
\label{fig:apigraphsense_overview}
\end{figure*}

\section{Approach}
\label{sec: approach}



\subsection{Overview}
\APIGH is a traffic-based API discovery framework composed of a denoising and normalization layer followed by a two-stage clustering layer, as illustrated in~\Cref{fig:apigraphsense_overview}, which incrementally abstracts raw HTTP traffic into API assets.
To improve robustness against traffic diversity and noise (Challenge~1), we first preprocess runtime traffic by filtering non-API requests and normalizing request paths before clustering.
To achieve better precision under fully unsupervised settings (Challenge~2), our approach avoids premature merging by first clustering requests at the interface level and applying semantic refinement only within each template group.

\subsection{Statistical Traffic Denoising and Path Normalization}

Raw HTTP traffic is noisy and heterogeneous. Besides genuine API calls, logs often contain large amounts of non-API requests such as static resources, web pages, and malformed entries, which can substantially distort downstream clustering. Moreover, even within API traffic, URLs frequently embed dynamic identifiers (e.g., user IDs, order IDs) that create high-variance paths and fragment otherwise identical interfaces. To address these issues, we preprocess traffic with (i) multi-signal noise filtering to remove non-API requests, and (ii) identifier pattern learning for path normalization, which abstracts dynamic segments into stable interface templates.

\textbf{Stage 1: Multi-signal noise filtering.}
We filter out non-API requests using a cascaded set of low-cost checks. Concretely, a request is removed if it matches any of the following signals: (1) static extensions (e.g., \texttt{.js}, \texttt{.css}, \texttt{.png}, \texttt{.svg}, \texttt{.html}), (2) static path patterns (e.g., \texttt{/static/}, \texttt{/assets/}, \texttt{/images/}) or common CDN/static-host signatures, and (3) the presence and value of the \texttt{Content-Type} header. Requests without a \texttt{Content-Type} field are discarded, and requests whose \texttt{Content-Type} indicates non-API semantics (e.g., \texttt{text/html}, \texttt{image/*}, \texttt{font/*}, media, archives) are pruned.
After rule-based filtering, we apply a lightweight logistic sanity check to remove extreme outliers. For each remaining request $r$, we compute:
\begin{equation}
p = \sigma(z) = \frac{1}{1 + e^{-z}},
\end{equation}
where $z$ is a weighted combination of simple structural features, including the HTTP method, path depth, whether the normalized path contains identifier placeholders, whether URL query parameters exist, and whether the payload type is JSON-based. Requests with $p < \tau$(threshold)$ = 0.01$ are removed.

\textbf{Stage 2: Path Normalization.}
After noise filtering, we normalize request paths into a canonical form to reduce superficial variations. Specifically, we (i) strip the scheme and host (e.g., http://, domain name), (ii) remove query strings and fragments, and (iii) standardize path formatting (e.g., collapse repeated slashes and trim trailing slashes when applicable). The resulting normalized path, together with the HTTP method, is used as the API interface representation for downstream clustering.

\subsection{Two-Stage Clustering with Structural Templates and Semantic Graphs.}
API traffic is highly heterogeneous.
Requests with identical URL structures may exhibit different operational behaviors, while directly clustering raw traffic based on semantic features is often unstable and expensive at scale.
To address this, we adopt a two-stage clustering framework that incrementally refines API requests from coarse structural patterns to fine-grained behavioral groups (Algorithm~\ref{alg:two_layer_apigraphsense}).
The first stage performs structure-level abstraction to group requests by interface shape, removing variability caused by dynamic path segments.
The second stage refines each structural group by separating distinct behaviors using lightweight semantic signals.
The two-stage clustering approach combines the robustness and efficiency of structural parsing with the expressiveness of semantic clustering, while avoiding unnecessary computation on noisy or sparse traffic.

\textbf{Stage 1: Template Mining (Structure perspective).}
The first stage aims to identify stable API \emph{structural templates} that represent interface shapes while removing variable path parts.
Given normalized request paths, each path is split into an ordered list of segments and inserted into a fixed-depth prefix tree using the Drain3~\cite{drain3_github}.
At each level of the tree, path segments are matched exactly whenever possible. Segment positions with many different values are replaced with wildcards.
This groups concrete request paths that differ only in identifier segments, such as \texttt{/api/v1/items/123} and \texttt{/api/v1/items/456}, into a single interface-level template \texttt{/api/v1/items/*}, as shown in~\Cref{fig:template_mining}.

\begin{figure}[htbp]
\centering
\scriptsize
\renewcommand{\arraystretch}{1.15}

\begin{tabular}{c c c}
\shortstack[l]{\texttt{/api/v1/items/123}\\
              \texttt{/api/v1/items/456}\\
              \texttt{/api/v1/items/789}}
& $\rightarrow$ &
\textbf{\texttt{/api/v1/items/*}}
\\[8pt]

\shortstack[l]{\texttt{/api/v1/order/812/status}\\
              \texttt{/api/v1/order/947/status}}
& $\rightarrow$ &
\textbf{\texttt{/api/v1/order/*/status}}
\end{tabular}

\caption{Structural template mining via Drain3. Concrete API paths are abstracted into interface-level templates by replacing dynamic identifier segments.}
\label{fig:template_mining}
\end{figure}

\textbf{Stage 2: Semantic refinement (Semantic perspective).}
The second stage aims to further separate requests within each structural template by considering their semantic differences.
Although requests grouped under the same template share the same path structure, they may still represent different behaviors.

\begin{table}[htbp]
\centering
\scriptsize
\setlength{\tabcolsep}{2mm}
\renewcommand\arraystretch{1.5}
\caption{Lightweight semantic features used to distinguish requests under the same structural template.}
\begin{tabular}{
p{20mm}   
p{30mm}   
p{25mm}   
}
\noalign{\hrule height 1.2pt}
\textbf{Feature Type} & \textbf{Examples} & \textbf{Purpose} \\
\noalign{\hrule height 1.2pt}
Path structure 
& path depth, API keywords (e.g., \texttt{api}, \texttt{v1}) 
& capture interface shape and action cues \\
Query parameters 
& parameter count, common keys (e.g., \texttt{page}, \texttt{limit}) 
& distinguish list-style and control behaviors \\
Payload complexity 
& body size, field count, nesting depth 
& separate simple Requests from complex traffic \\
\noalign{\hrule height 1.2pt}
\end{tabular}

\label{tab:semantic_features}
\end{table}

To distinguish requests under the same structural template, each request is represented by a lightweight semantic feature vector composed of simple path, query, and payload features in ~\Cref{tab:semantic_features}.
A semantic similarity graph is built within each template group using these feature representations.

Edges are added only between request pairs with high semantic similarity, filtering out weak or noisy connections.
Clustering is then framed as a refinement step that enforces consistent behavior within each graph.
Request embeddings are learned by minimizing the consistency loss in Equation~\eqref{eq:consistency}, which preserves strong semantic relations in the embedding space:
\begin{equation}
\label{eq:consistency}
\mathcal{L}_{\mathrm{cons}} = \left\| A - \sigma(ZZ^\top) \right\|_F^2 ,
\end{equation}
where $A$ is the semantic similarity matrix for a template group, $Z$ is the learned request embedding matrix, and $\sigma(\cdot)$ is the sigmoid function.

To produce discrete API behavior clusters, we add a clustering regularization term that encourages compact and stable groups in the embedding space.
The final semantic clustering objective is:
\begin{equation}
\label{eq:semantic_objective}
\mathcal{L} = \mathcal{L}_{\mathrm{cons}} + \lambda \mathcal{L}_{\mathrm{clus}},
\end{equation}
where $\mathcal{L}_{\mathrm{clus}}$ is the clustering regularization term, and $\lambda$ controls the trade-off between behavior consistency and cluster separation. Graph-based semantic clustering is used only when sufficient samples and dense similarity graphs are available; otherwise, we fall back to K-means on the feature vectors to ensure robustness.

\begin{algorithm}[htbp]
\scriptsize
\caption{Two-Stage Clustering with Structural Templates and Semantic Graphs}
\label{alg:two_layer_APIGraphSense }
\SetAlgoNlRelativeSize{-1}

\KwIn{Normalized API paths $\{\hat{p}_i\}$, request feature vectors $\{\mathbf{x}_i\}$}
\KwOut{Discovered API assets $\mathcal{C}$}

\BlankLine
\textbf{Stage 1: Template Mining}\;

Build prefix tree over $\{\hat{p}_i\}$ using Drain3\;
Assign each request to a template group $\mathcal{G}_k$\;

\BlankLine
\textbf{Stage 2: Semantic refinement}\tcp*{separate behaviors under same template}

$\mathcal{C} \leftarrow \emptyset$\;

\ForEach{template group $\mathcal{G}_k$}{
    \uIf{$|\mathcal{G}_k| < 3$}{
        $\mathcal{C} \leftarrow \mathcal{C} \cup \{\mathcal{G}_k\}$
        \textbf{continue}
    }

    Build similarity graph $A$ from $\{\mathbf{x}_i\}_{i\in\mathcal{G}_k}$\tcp*{semantic affinity}

    \uIf{\textbf{DAEGC is applicable on $A$}}{
        Train DAEGC on $(A, \{\mathbf{x}_i\})$\tcp*{split semantic behaviors}
        $\mathcal{C}_k \leftarrow$ cluster assignments from DAEGC\;
    }
    \Else{
        $\mathcal{C}_k \leftarrow$ KMeans$(\{\mathbf{x}_i\})$\tcp*{fallback}
    }

    $\mathcal{C} \leftarrow \mathcal{C} \cup \mathcal{C}_k$\;
}

\KwRet{$\mathcal{C}$}\;
\end{algorithm}

\section{Experiment Design}
\label{sec:experiment design}


\subsection{Research Questions}

To evaluate the proposed method from multiple perspectives, we formulate three research questions (RQs).

\textbf{RQ1. How effective is \APIGH in discovering APIs across different applications?}

\textbf{\textit{Motivation.}}
Effectiveness is a fundamental requirement of any API discovery approach.
In many practical scenarios, API traffic is collected from a single application with limited cross-project interference.
Under such settings, an API discovery method should accurately identify API structures with \emph{high precision}, avoiding incorrect endpoint merges, while maintaining robust performance across different applications.
Therefore, RQ1 evaluates the effectiveness of \APIGH on multiple representative web applications, with particular attention to its precision and cross-project robustness, and compares it against existing baseline methods.

\textbf{RQ2. How robust is \APIGH to noise in diverse API traffic?}

\textbf{\textit{Motivation.}}
Robustness is critical for deploying API discovery methods in practical analysis environments.
In real-world traffic, API requests often originate from multiple applications and are affected by various types and levels of noise.
Under such conditions, an API discovery method should maintain stable performance despite interference.
Therefore, RQ2 evaluates the robustness of \APIGH under mixed multi-project traffic with different noise types and noise levels, and compares its performance stability against baseline approaches.

\textbf{RQ3. How does each component contribute to the overall effectiveness of \APIGH?}

\textbf{\textit{Motivation.}}
Understanding the contribution of individual components is essential for validating the design of a complex API discovery system.
\APIGH adopts a multi-stage pipeline composed of several interacting components.
Under such a design, the overall effectiveness may depend on the joint impact of these components rather than any single module.
Therefore, RQ3 investigates the contribution of each component through ablation studies to assess its impact on the overall effectiveness of \APIGH.

\subsection{Studied Applications}
\label{sec:dataset}

To evaluate \APIGH under realistic multi-tenant \emph{web application} settings, we select six open-source web applications using explicit and systematic criteria.
Specifically, the selected projects must
(1) cover diverse functional domains to avoid application-specific bias;
(2) exhibit heterogeneous architectures, including both monolithic and microservice-based designs;
(3) expose a non-trivial set of structured HTTP(S) APIs that are exercised during normal user interactions; and
(4) be actively maintained and practically relevant, as reflected by community adoption and recent updates.

\begin{table}[htbp]
\centering
\scriptsize
\setlength{\tabcolsep}{0.6mm}
\renewcommand\arraystretch{1.5}
\caption{Basic characteristics of the selected web applications.
Reqs.: number of HTTP(S) requests collected;
Eps: number of unique API endpoints;
Upd.: last major update year.}
\label{tab:saas-characteristics}
\begin{tabular}{%
l
lll
p{2.6cm}
llr}
\noalign{\hrule height 1.2pt}
\textbf{Project} &
\textbf{Version} &
\textbf{Req.} &
\textbf{Eps} &
\textbf{Function} &
\textbf{Scale} &
\textbf{Stars} &
\textbf{Upd.} \\
\noalign{\hrule height 1.2pt}
Train-Ticket & 1.0.0 & 6564 & 13 & Ticket booking & Small & 2k & 2024 \\
Humhub       & 1.15.1 & 3957 & 18 & Social networking & Medium & 6.6k & 2025 \\
Memos        & 0.23.0 & 508  & 25 & Note taking & Medium & 47.3k & 2025 \\
Overleaf     & 5.2.1 & 162  & 32 & Collaborative editing & Large & 17k & 2025 \\
Nextcloud    & 32.0.3 & 107  & 20 & Cloud storage & Large & 33.6k & 2025 \\
Dify         & 1.9.0 & 1354 & 91 & LLM workflow platform & Large & 122k & 2026 \\
\noalign{\hrule height 1.2pt}
\end{tabular}
\end{table}

\begin{table*}[htbp]
\centering
\scriptsize
\caption{Atomic noise rules used for noisy dataset construction}
\label{tab:noise-rules}
\setlength{\tabcolsep}{4.5mm}
\renewcommand\arraystretch{1.15}
\begin{tabular}{p{1.5cm} ll}

\noalign{\hrule height 1.2pt}
\textbf{Noise Type} & \textbf{Rule} & \textbf{Example} \\
\noalign{\hrule height 1.2pt}
\multirow{14}{*}{Lexify}
& Query Order Shuffle
& \texttt{/api/user?id=1\&role=admin} $\rightarrow$ \texttt{/api/user?role=admin\&id=1} \\

& Neutral Query Parameter
& \texttt{/api/user?id=1} $\rightarrow$ \texttt{/api/user?id=1\&tmp=0} \\

& Duplicate Query Key
& \texttt{/api/user?id=1} $\rightarrow$ \texttt{/api/user?id=1\&id=1} \\

& Underscore Injection
& \texttt{/api/user/profile} $\rightarrow$ \texttt{/api/user\_/profile} \\

& Hyphen Duplication
& \texttt{/api/user-profile} $\rightarrow$ \texttt{/api/user--profile} \\

& Dot Injection
& \texttt{/api/user} $\rightarrow$ \texttt{/api/us.er} \\

& Repeated Slash
& \texttt{/api/user/profile} $\rightarrow$ \texttt{/api//user/profile} \\

& Trailing Slash Addition
& \texttt{/api/user} $\rightarrow$ \texttt{/api/user/} \\

& Trailing Slash Removal
& \texttt{/api/user/} $\rightarrow$ \texttt{/api/user} \\

& Uppercase Token
& \texttt{/api/user} $\rightarrow$ \texttt{/API/user} \\

& Lowercase Token
& \texttt{/API/User} $\rightarrow$ \texttt{/api/user} \\

& Space Encoding (\texttt{\%20})
& \texttt{/api/search?q=hello world} $\rightarrow$ \texttt{/api/search?q=hello\%20world} \\

& Plus Encoding (\texttt{+})
& \texttt{/api/search?q=hello world} $\rightarrow$ \texttt{/api/search?q=hello+world} \\

& Hex Encoding
& \texttt{/api/search?q=test} $\rightarrow$ \texttt{/api/search?q=\%74\%65\%73\%74} \\

\hline
\multirow{9}{*}{Interfere}
& Static Asset Request
& \texttt{/static/app.js}, \texttt{/assets/style.css} \\

& Image Resource Request
& \texttt{/images/logo.png}, \texttt{/img/banner.jpg} \\

& Font and Media Request
& \texttt{/fonts/main.woff2}, \texttt{/media/intro.mp4} \\

& Health Check Endpoint
& \texttt{/health}, \texttt{/status} \\

& Metrics Endpoint
& \texttt{/metrics}, \texttt{/actuator/metrics} \\

& Framework Handshake Request
& \texttt{/sockjs/info}, \texttt{/ws/connect} \\

& Hot Reload / Dev Channel
& \texttt{/webpack-hmr}, \texttt{/vite/client} \\

& Third-party Analytics Call
& \texttt{/analytics/collect}, \texttt{/track/event} \\

& CDN / Proxy Trace
& \texttt{/cdn-cgi/trace}, \texttt{/proxy/ping} \\

\noalign{\hrule height 1.2pt}
\end{tabular}
\end{table*}

Based on these criteria, we select six representative open-source web applications, as summarized in Table~\ref{tab:saas-characteristics}, covering collaborative editing, ticket booking, note taking, social networking, cloud storage, and LLM workflow orchestration.
The selected systems exhibit distinct functional roles and architectural styles.
Overleaf and Nextcloud represent large-scale collaborative web platforms with intensive and fine-grained API interactions;
Train-Ticket is a widely used microservice-based benchmark web system featuring complex inter-service API dependencies;
Memos exposes relatively clean and lightweight REST-style APIs typical of single-purpose web applications;
Humhub provides modular social networking functionalities with diverse user-driven API behaviors;
and Dify is a rapidly growing and actively evolving open-source web platform for LLM-based application development, introducing modern agent-oriented and workflow-driven API interaction patterns.

Each application is deployed locally and exercised through browser-driven interactions to emulate realistic user behaviors.
All HTTP(S) traffic is directly captured from Burp Suite Pro configured as an interception proxy, recording complete request--response pairs.
As reported in Table~\ref{tab:saas-characteristics}, Req. denotes the total number of captured HTTP(S) requests, while EPs indicates the number of unique API endpoints after normalization and deduplication.
To obtain ground-truth semantic labels for evaluation, API endpoints are manually annotated by analyzing the application source code and publicly available API documentation.
This labeling process required 32 person-hours of manual effort and provides reliable reference labels for assessing the accuracy of API graph construction.

\subsection{Noisy Simulation}
\label{sec:noisy-dataset}

To evaluate the robustness of API endpoint discovery under realistic operating conditions, we construct noisy variants of each single-project dataset by introducing controlled perturbations into the original traffic.
These perturbations reflect common artifacts of real-world HTTP-based systems, where API calls coexist with heterogeneous client behaviors and non-API traffic.

We consider two complementary noise types, termed \textbf{Lexify} and \textbf{Interfere}.
Lexify models lexical variability within individual API requests, while Interfere models structural interference caused by semantically irrelevant HTTP traffic.
Both noise types are applied in a semantics-preserving manner and are used solely for robustness evaluation.

\textbf{Lexify.}
Lexify introduces fine-grained syntactic variability within individual HTTP requests without altering the invoked API endpoint.
It simulates inconsistencies arising from different client implementations, serialization behaviors, and framework-level preprocessing by applying syntax-preserving transformations such as query parameter reordering, neutral parameter insertion, encoding and delimiter variations, case changes, and trailing-slash modifications.

\textbf{Interfere.}
Interfere introduces structural noise by injecting realistic but semantically irrelevant HTTP requests into the API traffic.
This models mixed-traffic deployment environments where API calls are interleaved with resource loading, framework-generated background requests, and auxiliary service interactions.
The injected requests introduce additional path patterns unrelated to the target API surface and are excluded from ground-truth endpoint annotations.

The complete set of noise rules used by Lexify and Interfere is summarized in Table~\ref{tab:noise-rules}.
For each single-project dataset, we generate multiple evaluation variants by applying Lexify and Interfere independently.
Unless otherwise specified, the injected noise does not modify the original ground-truth endpoint annotations and is used solely to assess robustness under lexical variability and structural distraction.

\subsection{Selected Baseline}

To evaluate API endpoint discovery, we compare our approach with a set of representative baselines that operate at runtime or on semi-structured artifacts.
These baselines are grouped into three categories according to their primary discovery mechanism:
\textbf{(1) open-source tools}, \textbf{(2) log-based template mining methods}, and \textbf{(3) API traffic-based endpoint discovery methods}.

\noindent\textbf{(1) Open-source tools} extract API endpoint information from runtime request--response traces or network traffic and are widely used for API reverse engineering and specification generation.

\begin{itemize}
  \item \textbf{Optic}~\cite{optic}.
  An open-source tool that captures real HTTP traffic to generate and validate OpenAPI specifications, enabling endpoint discovery from observed API interactions.

  \item \textbf{Mitmproxy2Swagger}~\cite{mitmproxy2swagger}.
  A reverse-engineering tool built on mitmproxy that automatically generates OpenAPI specifications from intercepted HTTP/HTTPS traffic, suitable for discovering undocumented or black-box APIs.
\end{itemize}

\noindent\textbf{(2) Log-based template mining methods} discover API endpoint patterns by mining templates from runtime traffic logs.
They extract invariant textual structures that represent endpoint templates, without relying on static specifications or source code.

\begin{itemize}
  \item \textbf{LogCluster}~\cite{vaarandi2015logcluster}.
  A data mining-based log clustering algorithm that groups similar log entries by identifying frequent token patterns and replacing variable tokens with wildcards.

  \item \textbf{LogNgram}~\cite{dai2022logram}.
  A template mining approach that uses token-level n-gram statistics to distinguish static components from dynamic variables in runtime logs.

  \item \textbf{UniParser}~\cite{liu2022uniparser}.
  A unified log parsing method that leverages neural encoders to learn common structural and semantic patterns across heterogeneous runtime logs.

  \item \textbf{LogPPT}~\cite{logppt}.
  A prompt-based log template mining approach that employs pretrained language models to extract templates from runtime logs.
  We consider both the RoBERTa-based and GPT-based variants under few-shot settings.
\end{itemize}

\noindent\textbf{(3) API traffic-based endpoint discovery methods} explicitly target API endpoint discovery from runtime observations, documentation, or interaction traces.
They are designed to infer API endpoints, path templates, or specifications, rather than general-purpose log templates.

\begin{itemize}
  \item \textbf{Web API Search}~\cite{liu2020webapisearch}.
  A learning-based approach that supports endpoint-level API discovery by matching natural language queries against API and endpoint descriptions mined from online documentation.

  \item \textbf{APID2Spec}~\cite{yang2018extracting}.
  A documentation-driven approach that extracts API specifications by crawling documentation pages and inferring base URLs, path templates, and HTTP methods.

  \item \textbf{APICARV}~\cite{yandrapally2023apicarv}.
  A dynamic analysis technique that infers API endpoints by monitoring runtime API traffic generated through UI interactions and constructing an API graph from observed traces.

  \item \textbf{APIDrain3}.
  An online template mining approach adapted to API traffic, which incrementally extracts URL templates from streaming API flows for continuous endpoint discovery.
\end{itemize}

\subsection{Evaluation Metrics}

We evaluate the correctness of API endpoint discovery and the structural quality of clustering using multiple quantitative metrics, covering both endpoint-level accuracy and clustering consistency.

To evaluate the accuracy of discovered API endpoints at the set level, we use Precision of Group Accuracy (PGA), Recall of Group Accuracy (RGA), and the F1-score of Group Accuracy ($\textbf{FGA}$).
Precision measures the proportion of discovered endpoints that correspond to ground-truth endpoints, Recall measures the coverage of ground-truth endpoints, and F1-score provides a balanced summary of the two:
\begin{equation}
\small
\begin{gathered}
\textbf{PGA} = \frac{TP}{TP + FP}, \qquad
\textbf{RGA} = \frac{TP}{TP + FN}, \\
\textbf{FGA} = \frac{2TP}{2TP + FP + FN}.
\end{gathered}
\end{equation}

where $TP$ denotes the number of correctly discovered endpoints, $FP$ the number of spurious endpoints, and $FN$ the number of ground-truth endpoints that are not recovered.

To evaluate whether each predicted cluster corresponds to a single semantic API endpoint, we measure cluster purity(Purity).
Let $\{x_i\}_{i=1}^{N}$ denote the set of API endpoint templates, where each template $x_i$ is associated with a ground-truth label $y_i$ and a predicted cluster label $\hat{y}_i$.
Cluster purity is defined as:
\begin{equation}
\small
\text{Purity} = \frac{1}{N} \sum_k \max_{c} \left| \{ x_i : \hat{y}_i = k \wedge y_i = c \} \right|,
\end{equation}
where $k$ indexes predicted clusters and $c$ indexes ground-truth endpoint labels.

Together, these metrics provide a unified evaluation of endpoint discovery accuracy and clustering structure using a consistent notation across all metrics.

\section{Result Analysis}
\label{sec: Result analysis}

\subsection{Effectiveness (RQ1)}
\label{sec:rq1-results}

To evaluate the ability of \APIGH to handle diverse projects with different scales and functionalities, we conduct experiments on six projects. The results show that \APIGH achieves strong and stable performance across all projects, demonstrating its effectiveness and robustness under diverse project settings.

\begin{table*}[htbp]
\centering
\scriptsize
\setlength{\tabcolsep}{1.3mm}
\renewcommand\arraystretch{1.5}
\caption{Performance of API Discovery on different projects (PGA, RGA, FGA in \%)}
\label{tab:rq1-results}
\begin{tabular}{l|ccc|ccc|ccc|ccc|ccc|ccc}
\noalign{\hrule height 1.2pt}
\multirow{2}{*}{\textbf{Method}} &
\multicolumn{3}{c|}{\textbf{Train-Ticket}} &
\multicolumn{3}{c|}{\textbf{Humhub}} &
\multicolumn{3}{c|}{\textbf{Memos}} &
\multicolumn{3}{c|}{\textbf{Overleaf}} &
\multicolumn{3}{c|}{\textbf{Nextcloud}} &
\multicolumn{3}{c}{\textbf{Dify}} \\
& \textbf{PGA} & \textbf{RGA} & \textbf{FGA}
  & \textbf{PGA} & \textbf{RGA} & \textbf{FGA}
  & \textbf{PGA} & \textbf{RGA} & \textbf{FGA}
  & \textbf{PGA} & \textbf{RGA} & \textbf{FGA}
  & \textbf{PGA} & \textbf{RGA} & \textbf{FGA}
  & \textbf{PGA} & \textbf{RGA} & \textbf{FGA} \\
\noalign{\hrule height 1.0pt}

Optic
& 69.23 & 84.62 & 76.15
& 22.22 & 11.11 & 14.81
& 54.17 & 100.00 & 70.27
& 43.75 & 37.50 & 40.38
& 33.33 & 15.01 & 20.69
& 25.00 & 16.48 & 19.87 \\

Mitproxy2Swagger
& 8.28 & 100.00 & 15.29
& 75.00 & 100.00 & 85.71
& 55.56 & 100.00 & 71.43
& 38.55 & 100.00 & 55.65
& 54.05 & 100.00 & 70.18
& 40.64 & 97.80 & 57.42 \\

\hline
LogCluster
& 86.67 & 100.00 & 92.86
& 90.00 & 100.00 & 94.74
& 78.57 & 88.00 & 83.02
& 34.78 & 25.00 & 29.09
& 77.78 & 35.00 & 48.28
& 71.43 & 60.44 & 65.48 \\

LogNgram
& 86.67 & 100.00 & 92.86
& 64.29 & 100.00 & 78.26
& 75.76 & 100.00 & 86.21
& 46.27 & 96.88 & 62.63
& 82.61 & 95.00 & 88.37
& 67.42 & 97.80 & 79.82 \\

UniParser
& 86.67 & 100.00 & 92.86
& 64.29 & 100.00 & 78.26
& 74.19 & 92.00 & 82.14
& 58.00 & 90.62 & 70.73
& 82.61 & 95.00 & 88.37
& 67.18 & 96.70 & 79.28 \\

LogPPT
& 86.67 & 100.00 & 92.86
& 64.29 & 100.00 & 78.26
& 75.76 & 100.00 & 86.21
& 46.27 & 96.88 & 62.63
& 82.61 & 95.00 & 88.37
& 67.42 & 97.80 & 79.82 \\

LogPPT (GPT5)
& 26.67 & 30.77 & 28.57
& 64.29 & 100.00 & 78.26
& 25.00 & 32.00 & 28.07
& 32.69 & 53.12 & 40.48
& 39.13 & 45.00 & 41.86
& 40.00 & 57.14 & 47.06 \\

\hline
WebAPISearch
& 80.00 & 30.77 & 44.44
& 69.23 & 50.00 & 58.06
& 50.00 & 16.00 & 24.24
& 55.00 & 34.38 & 42.31
& 76.92 & 50.00 & 60.61
& 27.78 & 10.99 & 15.75 \\

APID2Spec
& 86.67 & 100.00 & 92.86
& 64.29 & 100.00 & 78.26
& 72.73 & 64.00 & 68.09
& 50.00 & 90.62 & 64.44
& 81.82 & 90.00 & 85.71
& 65.69 & 73.63 & 69.43 \\

APICARV
& 86.67 & 100.00 & 92.86
& 64.29 & 100.00 & 78.26
& 75.76 & 100.00 & 86.21
& 46.27 & 96.88 & 62.63
& 82.61 & 95.00 & 88.37
& 67.42 & 97.80 & 79.82 \\

APIDrain3
& 86.67 & 100.00 & 92.86
& 81.82 & 100.00 & 90.00
& 74.19 & 92.00 & 82.14
& 62.50 & 62.50 & 62.50
& 78.95 & 75.00 & 76.92
& 67.42 & 97.80 & 79.82 \\

\hline
\APIGH
& \textbf{100.00} & \textbf{84.62} & \textbf{91.67}
& \textbf{85.71} & \textbf{100.00} & \textbf{92.31}
& \textbf{96.15} & \textbf{100.00} & \textbf{98.04}
& \textbf{100.00} & \textbf{93.75} & \textbf{96.77}
& \textbf{94.74} & \textbf{90.00} & \textbf{92.31}
& \textbf{98.89} & \textbf{97.80} & \textbf{98.34} \\

\noalign{\hrule height 1.2pt}
\end{tabular}
\end{table*}

\textbf{\APIGH achieves the best precision (PGA) across diverse projects.}
As shown in~\Cref{tab:rq1-results}, on the simple Train-Ticket project, \APIGH achieves a PGA of 100\% and an FGA of 91.67\%, while maintaining an RGA of 84.62\%, thereby avoiding the over-merging behavior of baselines that reach RGA = 100\% at the cost of precision.
On medium-scale projects such as Humhub and Memos, \APIGH maintains high precision, achieving PGA values of 85.71\% and 96.15\%, respectively, while both projects reach an RGA of 100\%, effectively suppressing incorrect endpoint merges.
On large and complex projects such as Overleaf and Dify, \APIGH continues to deliver very high precision, with PGA values of 100\% and 98.89\%, demonstrating strong robustness as project scale and API complexity increase.

\textbf{\APIGH consistently achieves the best overall performance (FGA) across projects of different scales.}
It achieves the highest FGA on all six projects, with particularly large improvements over the strongest baselines on Overleaf (96.77\%, +26.04), Memos (98.04\%, +11.83), and Dify (98.34\%, +18.52).
On medium-scale projects, \APIGH achieves FGA values of 92.31\% on Humhub and 98.04\% on Memos, benefiting from its balanced precision and recall.
On large and complex projects, it substantially outperforms all baseline methods, reaching an FGA of 96.77\% on Overleaf, 92.31\% on Nextcloud, and 98.34\% on Dify, demonstrating strong scalability and robustness to complex API semantics.

\begin{figure}[t]
    \centering
    \includegraphics[width=1.0\columnwidth]{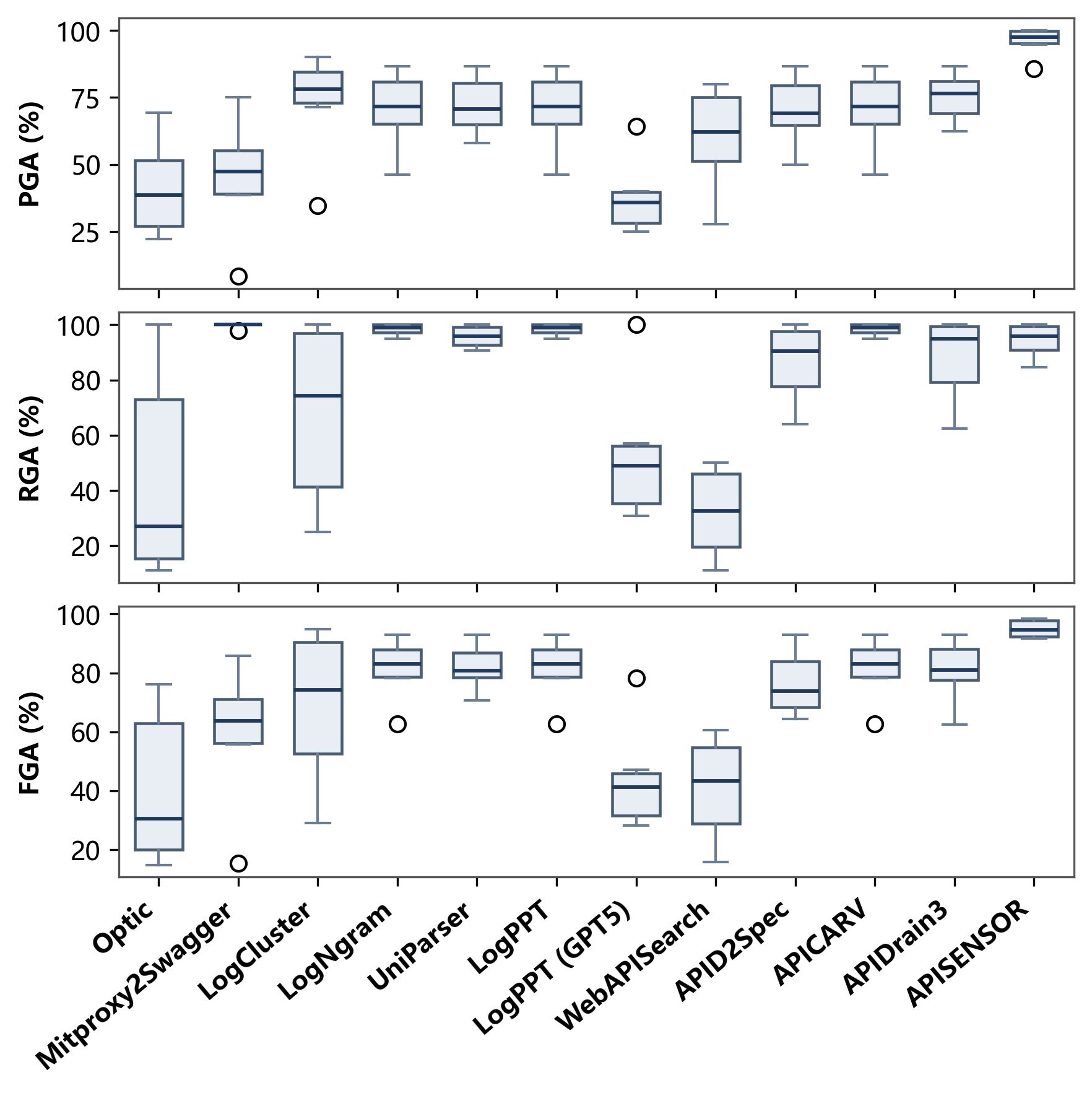}
    \caption{Performance Stability of API Endpoint Discovery Methods Across Diverse Projects (PGA, RGA, FGA in \%)}
    \label{fig:api_metrics_boxplot_sharedx}
\end{figure}

\textbf{\APIGH demonstrates the most stable high-performance on both PGA and FGA across diverse projects.}
As shown in Figure~\ref{fig:api_metrics_boxplot_sharedx}, the PGA and FGA values of \APIGH are tightly clustered near the upper bound with small interquartile ranges and few outliers, indicating consistently strong performance rather than occasional peaks. Quantitatively, \APIGH achieves an average PGA of 95.92\% and an average FGA of 94.91\% across six projects, with low variances of 24.63 and 8.17, respectively. This shows that the high precision and overall accuracy of \APIGH are maintained consistently across projects of different scales and functionalities, demonstrating strong robustness to project diversity.

\begin{tcolorbox}[width=\linewidth, boxrule=1pt, sharp corners=all,
left=2pt, right=2pt, top=2pt, bottom=2pt, colback=white, colframe=black]
\textbf{Answer to RQ1:}
\APIGH achieves the best API endpoint discovery performance across all projects.
It consistently attains the highest precision, resulting in the fewest false-positive endpoint merges, with an average PGA of 95.92\%.
Meanwhile, it achieves the highest overall accuracy on every project, with an average FGA of 94.91\%, while remaining the most stable across diverse projects.
\end{tcolorbox}

\begin{table*}[t]
\centering
\scriptsize
\setlength{\tabcolsep}{1.1mm}
\renewcommand\arraystretch{1.5}
\caption{Single-project API endpoint discovery performance}
\label{tab:rq1-results}
\begin{tabular}{l|cccc|cccc|cccc}
\noalign{\hrule height 1.2pt}
\multirow{2}{*}{\textbf{Method}} &
\multicolumn{4}{c|}{\textbf{Standard}} &
\multicolumn{4}{c|}{\textbf{Interfere}} &
\multicolumn{4}{c}{\textbf{Lexify}} \\
& \textbf{PGA} & \textbf{RGA} & \textbf{FGA} & \textbf{Purity}
& \textbf{PGA} & \textbf{RGA} & \textbf{FGA} & \textbf{Purity}
& \textbf{PGA} & \textbf{RGA} & \textbf{FGA} & \textbf{Purity} \\
\noalign{\hrule height 1.0pt}

Optic
& 24.00 & 38.89 & 29.68 & 4.81
& 17.78($\downarrow$6.22) & 38.89($-$) & 24.40($\downarrow$5.28) & 4.81($-$)
& 18.38($\downarrow$5.62) & 39.39($\uparrow$0.50) & 25.07($\downarrow$4.61) & 4.81($-$) \\

Mitproxy2Swagger
& 24.20 & 19.19 & 21.41 & 3.80
& 7.58($\downarrow$16.62) & 7.58($\downarrow$11.61) & 7.58($\downarrow$13.83) & 3.04($\downarrow$0.76)
& 6.85($\downarrow$17.35) & 7.58($\downarrow$11.61) & 7.19($\downarrow$14.22) & 3.04($\downarrow$0.76) \\

\hline

LogCluster
& 73.65 & 62.12 & 67.40 & 90.06
& 57.21($\downarrow$16.44) & 62.12($-$) & 59.56($\downarrow$7.84) & 90.56($\uparrow$0.50)
& 66.85($\downarrow$6.80) & 62.12($-$) & 64.40($\downarrow$3.00) & 89.06($\downarrow$1.00) \\

LogNgram
& 68.18 & 98.48 & 80.58 & 90.68
& 54.47($\downarrow$13.71) & 98.48($-$) & 70.14($\downarrow$10.44) & 91.85($\uparrow$1.17)
& 50.92($\downarrow$17.26) & 97.47($\downarrow$1.01) & 66.90($\downarrow$13.68) & 89.89($\downarrow$0.79) \\

UniParser
& 71.43 & 95.94 & 81.90 & 90.45
& 56.51($\downarrow$14.92) & 96.46($\uparrow$0.52) & 71.27($\downarrow$10.63) & 91.55($\uparrow$1.10)
& 55.29($\downarrow$16.14) & 94.95($\downarrow$0.99) & 69.89($\downarrow$12.01) & 89.55($\downarrow$0.90) \\

LogPPT
& 68.18 & 98.48 & 80.58 & 90.68
& 54.47($\downarrow$13.71) & 98.48($-$) & 70.14($\downarrow$10.44) & 91.85($\uparrow$1.17)
& 51.88($\downarrow$16.30) & 97.47($\downarrow$1.01) & 67.72($\downarrow$12.86) & 89.95($\downarrow$0.73) \\

LogPPT (GPT5)
& 43.40 & 58.08 & 49.68 & 90.34
& 33.14($\downarrow$10.26) & 58.08($-$) & 42.20($\downarrow$7.48) & 91.55($\uparrow$1.21)
& 33.24($\downarrow$10.16) & 57.07($\downarrow$1.01) & 42.01($\downarrow$7.67) & 89.54($\downarrow$0.80) \\

\hline

WebAPISearch
& 51.76 & 22.22 & 31.10 & 48.36
& 33.33($\downarrow$18.43) & 22.22($-$) & 26.67($\downarrow$4.43) & 51.93($\uparrow$3.57)
& 48.89($\downarrow$2.87) & 22.22($-$) & 30.56($\downarrow$0.54) & 48.99($\uparrow$0.63) \\

APID2Spec
& 67.66 & 80.30 & 73.44 & 86.34
& 50.81($\downarrow$16.85) & 79.29($\downarrow$1.01) & 61.93($\downarrow$11.51) & 88.66($\uparrow$2.32)
& 48.58($\downarrow$19.08) & 77.78($\downarrow$2.52) & 59.81($\downarrow$13.63) & 85.50($\downarrow$0.84) \\

APICARV
& 68.18 & 98.48 & 80.58 & 90.68
& 54.47($\downarrow$13.71) & 98.48($-$) & 70.14($\downarrow$10.44) & 91.85($\uparrow$1.17)
& 51.88($\downarrow$16.30) & 97.47($\downarrow$1.01) & 67.72($\downarrow$12.86) & 89.89($\downarrow$0.79) \\

APIDrain3
& 72.24 & 89.39 & 79.91 & 90.16
& 56.15($\downarrow$16.09) & 89.90($\uparrow$0.51) & 69.13($\downarrow$10.78) & 91.00($\uparrow$0.84)
& 55.17($\downarrow$17.07) & 88.89($\downarrow$0.50) & 68.09($\downarrow$11.82) & 89.25($\downarrow$0.91) \\

\hline

\APIGH
& \textbf{98.25} & \textbf{84.85} & \textbf{91.06} & \textbf{91.24}
& \textbf{97.66($\downarrow$0.59)} & \textbf{84.34($\downarrow$0.51)} & \textbf{90.51($\downarrow$0.55)} & \textbf{91.72($\uparrow$0.48)}
& \textbf{83.59($\downarrow$14.66)} & \textbf{82.32($\downarrow$2.53)} & \textbf{82.95($\downarrow$8.11)} & \textbf{92.16($\uparrow$0.92)} \\

\noalign{\hrule height 1.2pt}
\end{tabular}
\end{table*}

\subsection{Robustness (RQ2)}
\label{sec:rq2-results}

To evaluate the robustness of \APIGH under noisy and mixed API traffic, we conduct experiments with increasing levels of cross-application interference and lexical noise.
The results show that \APIGH consistently maintains superior performance under all noise settings, demonstrating strong robustness to interference in diverse and realistic API traffic.

\textbf{Under the Standard setting, where API traffic from multiple projects is collected simultaneously at the gateway side, \APIGH consistently outperforms all baselines across all evaluation metrics.}
As shown in Table~\ref{tab:rq1-results}, \APIGH achieves a PGA of 98.25, an RGA of 84.85, and an FGA of 91.06, while also maintaining high clustering purity (91.24).
In contrast, baseline methods typically exhibit a clear trade-off between precision and recall, achieving very high RGA (e.g., up to 98.48) at the cost of substantially lower PGA (around 68.18), indicating aggressive and error-prone endpoint merging.
These results show that \APIGH can accurately distinguish endpoints across multiple coexisting projects without relying on over-merging strategies.

\begin{figure}[htbp]
    \centering
    \includegraphics[width=1.0\linewidth]{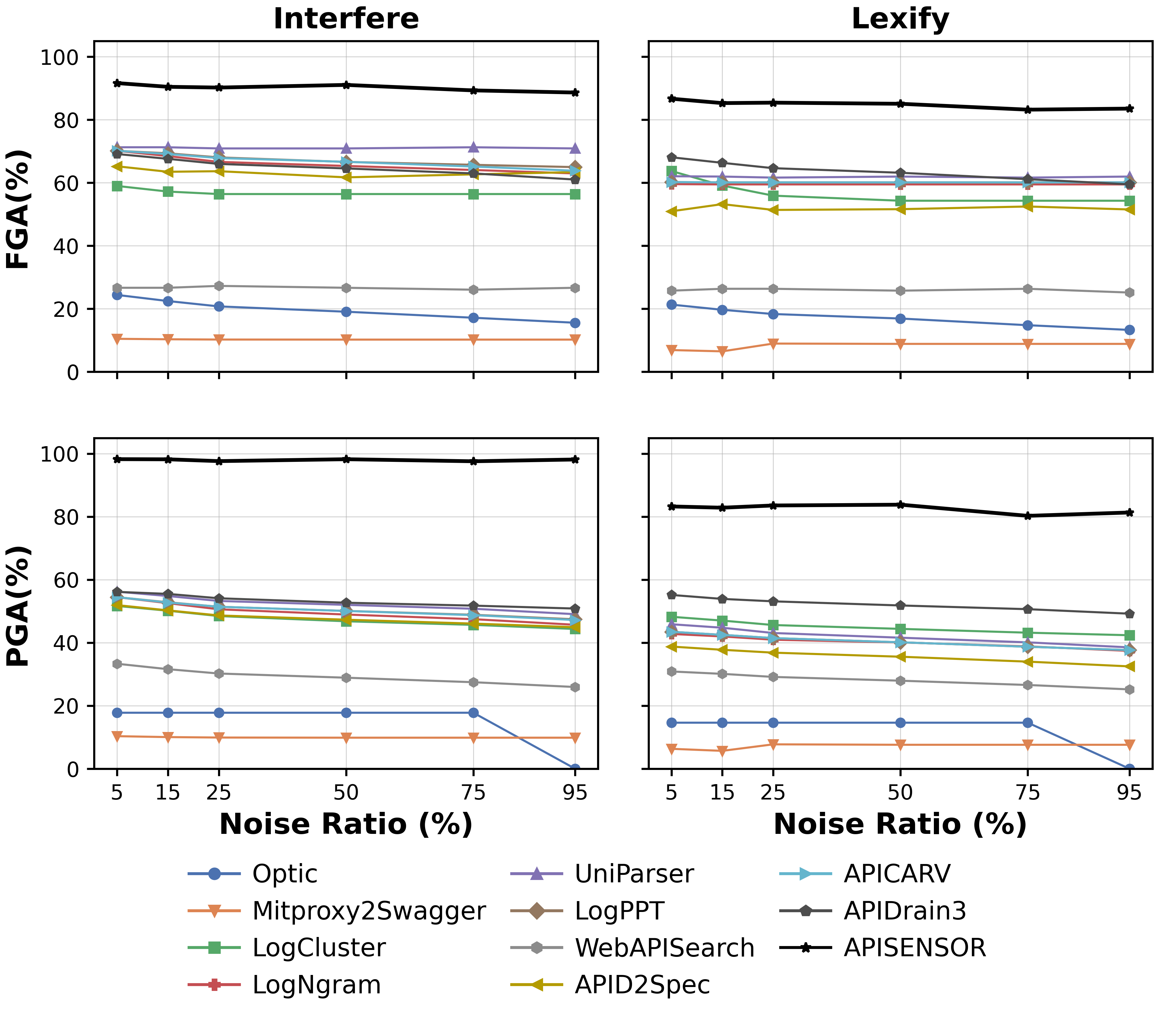}
    \vspace{-1.0ex}
    \caption{Robustness of API discovery under increasing noise ratios (Interfere and Lexify).}

    \label{fig:noise_robustness}
\end{figure}

\textbf{When cross-application interference is introduced, \APIGH is affected the least and maintains the highest performance.}
Under the Interfere setting, baseline methods suffer noticeable degradation, with FGA dropping to 59.56--71.27 due to confusion caused by interleaved traffic patterns.
In contrast, \APIGH maintains a PGA of 97.66 and an RGA of 84.34, resulting in an FGA of 90.51, with only a marginal decrease of 0.55 compared to the Standard setting.
This demonstrates that \APIGH remains highly robust to cross-application noise when deployed in realistic gateway environments.

\textbf{Even under the most challenging Lexify setting, \APIGH continues to deliver the strongest performance with the smallest degradation.}
While baseline methods degrade substantially when lexical cues are weakened, achieving at most an FGA of 69.89, \APIGH still attains a PGA of 83.59 and an RGA of 82.32, yielding an FGA of 82.95.
Despite the increased difficulty, \APIGH preserves a clear performance margin over all baselines, indicating strong generalization ability beyond surface-level lexical similarity in highly heterogeneous, multi-application traffic.

\textbf{\APIGH remains the most robust method as noise increases.}
As shown in Figure~\ref{fig:noise_robustness}, when the noise ratio increases under both the Interfere and Lexify settings, \APIGH consistently maintains the highest FGA and Purity among all methods. While the performance of baseline approaches drops noticeably as noise intensifies, \APIGH degrades only slightly. Specifically, its FGA decreases by less than 4\% as the noise ratio increases from 5\% to 95\%. Even at very high noise levels (e.g., 75\% and 95\%), \APIGH still preserves a clear performance margin over all baselines. These results demonstrate that \APIGH is highly resilient to cross-application interference and lexical corruption, maintaining stable API discovery performance under increasingly noisy traffic.

\begin{tcolorbox}[width=\linewidth, boxrule=1pt, sharp corners=all,
left=2pt, right=2pt, top=2pt, bottom=2pt, colback=white, colframe=black]
\textbf{Answer to RQ2:}
\APIGH consistently maintains the highest API endpoint discovery performance under all noisy settings, even in the presence of cross-application interference and weakened lexical cues.
Moreover, as the noise ratio increases, \APIGH exhibits the slowest performance degradation among all methods, demonstrating the strongest robustness to increasing noise in diverse and realistic API traffic.
\end{tcolorbox}
\begin{figure*}[t]
\centering
\includegraphics[width=0.85\linewidth]{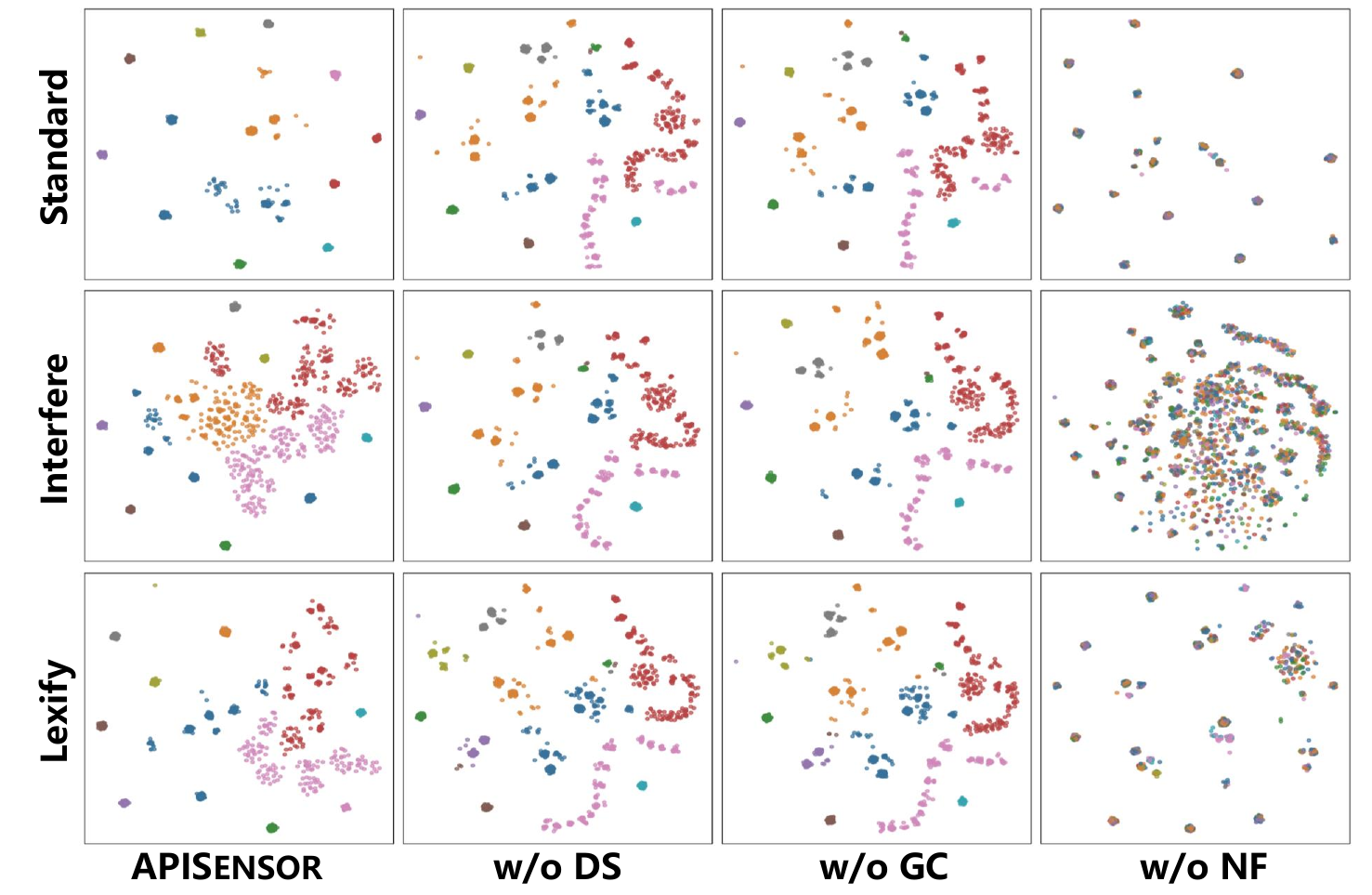}
\caption{The t-SNE visualizations of API endpoint representations under different modeling configurations and noise conditions.}
\label{fig: RQ3_Vision}
\end{figure*}

\subsection{Ablation Study (RQ3)}

To quantify the contribution of each component in \APIGH, we conduct a comprehensive ablation study under all three evaluation settings: \emph{Standard}, \emph{Interfere}, and \emph{Lexify}.
This design allows us to examine not only the standalone effectiveness of each component, but also its role under increasing levels of structural interference and lexical variability.

\APIGH consists of three key components:
(1) \emph{NF}, which normalizes raw HTTP requests to suppress instance-level noise;
(2) \emph{Structural Template Mining}, implemented using Drain3 to extract canonical API templates;
and (3) \emph{Graph-based Representation Learning}, implemented via the DAEGC module to jointly model structural and semantic relationships.
We construct three ablated variants by removing one component at a time while keeping the rest of the pipeline unchanged.

\begin{table}[htbp]
\centering
\scriptsize
\setlength{\tabcolsep}{0.3mm}
\renewcommand\arraystretch{1.15}
\caption{Ablation results of \APIGH.}
\label{tab:ablation}
\begin{tabular}{ll|cccc}
\noalign{\hrule height 1.2pt}
\multicolumn{2}{c|}{\textbf{Set.}} 
& \textbf{PGA} & \textbf{RGA} & \textbf{FGA} & \textbf{Purity} \\
\noalign{\hrule height 1.2pt}

\multirow{4}{*}{\textbf{Standard}} & \APIGH 
& 98.28 & 86.36 & 91.94 & 93.17 \\
 & w/o NF 
& 32.60($\downarrow$65.68) & 82.32($\downarrow$4.04) & 46.70($\downarrow$45.24) & 89.00($\downarrow$4.17) \\
 & w/o DS 
& 92.11($\downarrow$6.17) & 17.68($\downarrow$68.68) & 29.66($\downarrow$62.28) & 37.37($\downarrow$55.80) \\
 & w/o GC 
& 98.58($\uparrow$0.30) & 70.20($\downarrow$16.16) & 82.01($\downarrow$9.93) & 85.66($\downarrow$7.51) \\
\hline

\multirow{4}{*}{\textbf{Interfere}} & \APIGH 
& 97.69 & 85.35 & 91.11 & 92.04 \\
 & w/o NF 
& 25.35($\downarrow$72.34) & 63.64($\downarrow$21.71) & 36.26($\downarrow$54.85) & 76.60($\downarrow$15.44) \\       
 & w/o DS 
& 86.84($\downarrow$10.85) & 16.67($\downarrow$68.68) & 27.97($\downarrow$63.14) & 37.37($\downarrow$54.67) \\
 & w/o GC 
& 97.86($\uparrow$0.17) & 69.19($\downarrow$16.16) & 81.07($\downarrow$10.04) & 85.39($\downarrow$6.65) \\
\hline

\multirow{4}{*}{\textbf{Lexify}} & \APIGH 
& 83.76 & 83.33 & 83.54 & 92.56 \\
 & w/o NF 
& 29.59($\downarrow$54.17) & 79.80($\downarrow$3.53) & 43.17($\downarrow$40.37) & 87.41($\downarrow$5.15) \\
 & w/o DS 
& 92.11($\uparrow$8.35) & 17.68($\downarrow$65.65) & 29.66($\downarrow$53.88) & 36.78($\downarrow$55.78) \\
 & w/o GC 
& 80.65($\downarrow$3.11) & 63.13($\downarrow$20.20) & 70.82($\downarrow$12.72) & 87.96($\downarrow$4.60) \\
\noalign{\hrule height 1.2pt}
\end{tabular}

\vspace{1mm}
\raggedright
\footnotesize{\textit{Note:}
w/o NF removes URL normalization and request canonicalization;
w/o Drain3 removes structural template mining;
w/o Graph replaces graph-based deep clustering with K-means.}
\end{table}

\textbf{Effect of Noise Filtering (NF).}
As shown in~\Cref{tab:ablation}, removing NF causes substantial performance degradation across all settings, mainly affecting precision-related metrics.
Under \emph{Standard}, PGA drops from 98.28 to 32.60 ($\downarrow$65.68) and FGA from 91.94 to 46.70 ($\downarrow$45.24), while RGA only slightly decreases by 4.04.
Similar trends are observed under \emph{Interfere}, where PGA and FGA decrease by 72.34 and 54.85, respectively.
The impact is most pronounced under \emph{Lexify}, with PGA dropping by 54.17 and FGA by 40.37, indicating that normalization is crucial for mitigating lexical noise and preserving cluster precision.

\textbf{Effect of Structural Template Mining (Drain3).}
Removing structural template mining leads to severe recall collapse and purity degradation across all settings.
In the \emph{Standard} setting, RGA drops sharply from 86.36 to 17.68 ($\downarrow$68.68) and Purity from 93.17 to 37.37 ($\downarrow$55.80).
A similar pattern holds under \emph{Interfere} and \emph{Lexify}, where RGA consistently falls by over 65 points.
The results in~\Cref{tab:ablation} show that template abstraction provides essential structural guidance for distinguishing API endpoints, especially in noisy environments.

\textbf{Effect of Graph-based Representation Learning (DAEGC).}
As shown in~\Cref{tab:ablation}, replacing DAEGC with K-means mainly affects recall and overall clustering quality.
Under \emph{Standard}, RGA decreases from 86.36 to 70.20 ($\downarrow$16.16) and FGA from 91.94 to 82.01 ($\downarrow$9.93), while PGA slightly increases.
Comparable degradations are observed under \emph{Interfere} and \emph{Lexify}, with FGA drops of 10.04 and 12.72, respectively.
This indicates that graph-based joint optimization is critical for maintaining coherent clustering under interference and cross-project scenarios.

Figure~\ref{fig: RQ3_Vision} provides a qualitative analysis of learned API endpoint representations, showing that \APIGH consistently achieves strong endpoint-level discriminability, while representation quality degrades in a structured and interpretable manner as modeling components are removed and noise is introduced.

\textbf{In standard conditions, the full pipeline produces compact and well-separated clusters, with each endpoint occupying a distinct region in the embedding space.}
The first row in~\Cref{fig: RQ3_Vision} indicates that the learned representations can reliably distinguish endpoints even across multiple projects.
When individual components are ablated, clusters become less compact and their boundaries more fragmented, suggesting that strong discriminability already relies on the joint contribution of lexical, structural, and relational modeling.

\textbf{As semantic interference is introduced, the full model largely preserves inter-endpoint separation, whereas ablated configurations exhibit increasing overlap between clusters.}
Interfering samples blur the boundaries between endpoints in the embedding space as shown in~\Cref{fig: RQ3_Vision}, indicating that higher-level structural abstraction and graph-based relational learning are essential for preventing semantic noise from collapsing global discriminability.

\textbf{Under lexical perturbations, representation degradation follows a different pattern.}
Endpoint identities are not entirely lost; samples associated with the same endpoint still form locally compact groups in the~\Cref{fig: RQ3_Vision}.
However, these groups become spatially fragmented and lose consistent relative positioning, resulting in an embedding space with weak global geometric coherence.
As a consequence, discriminability becomes highly localized and unstable, demonstrating that effective lexical normalization is a prerequisite for maintaining coherent and globally discriminative endpoint representations.

\begin{tcolorbox}[width=\linewidth, boxrule=1pt, sharp corners=all,
left=2pt, right=2pt, top=2pt, bottom=2pt, colback=white, colframe=black]
\textbf{Answer to RQ3:}
The core components of \APIGH are all indispensable.
Removing any component leads to clear degradation in endpoint clustering, not only in overall performance but also in the stability and sharpness of clustering boundaries, as further evidenced by the representation visualizations.
Together, these components are necessary to produce coherent, well-separated, and robust API endpoint representations under noise.
\end{tcolorbox}

\section{Discussion}
\label{sec: Discussion}

\subsection{Shadow APIs Revealed by Runtime Traffic: Evidence from the Case Study of Dify}

During the evaluation of \APIGH, we observed a clear inconsistency between the official API documentation of Dify and the API endpoints exercised at runtime. A number of API endpoints repeatedly appeared in network traffic traces but were missing from the API Reference at the time of our study. These endpoints are callable, actively used by the system, and essential for Console-level functionalities, yet remain undocumented or only indirectly described. Such undocumented but operational interfaces constitute a class of \emph{shadow APIs} that are invisible to documentation-centric API discovery approaches.

Representative examples of these shadow APIs span different functional modules. In the Datasets module, endpoints such as
\texttt{/console/api/datasets/\{id\}/metadata} and
\texttt{/console/api/datasets/\{id\}/error-docs}
were frequently observed during dataset management operations but absent from the API Reference. In the Apps module, we identified undocumented endpoints including
\texttt{/console/api/apps/\{id\}/api-keys}, which is essential for application integration and key management, and
\texttt{/console/api/apps/\{id\}/advanced-chat/workf\\low-runs}, which supports workflow execution monitoring. These endpoints represent core application functionalities, yet they are not discoverable by developers or agents relying solely on the documented API surface.

To validate that these findings reflected genuine documentation gaps rather than artifacts of traffic analysis, we reported the discovered inconsistencies to the Dify developer community through an official documentation issue. A core collaborator acknowledged that, due to legacy reasons and rapid system evolution, the API reference had become out of sync with the implementation and contained notable gaps. The maintainers further indicated ongoing efforts to consolidate the API Reference as the single source of truth. This independent confirmation substantiates that the identified endpoints correspond to real and practically relevant shadow APIs.

This case study illustrates how shadow APIs naturally emerge in evolving web applications, even in actively maintained open-source projects. From a software engineering perspective, the presence of such shadow APIs complicates reuse, automation, and reliable integration. Developers and LLM-based agents relying solely on official API references may incorrectly assume that certain functionalities do not exist. The Dify case demonstrates that black-box, traffic-based API discovery is essential for reconstructing the effective API surface of real-world systems. By grounding API discovery in observed runtime behavior, \APIGH enables systematic identification of shadow APIs and provides a more faithful representation of application functionality than documentation-based approaches alone.

\subsection{Implications for LLM-Based and Tool-Using Agents}

The emergence of sophisticated autonomous agents such as Clawbot (also known as Moltbot or OpenClaw) underscores the transformative potential of LLM-based and tool-using agents in software ecosystems and end-user workflows. Unlike traditional chatbots that primarily provide text responses, agents like Clawbot integrate deeply with existing applications and infrastructure to perform concrete actions, such as interacting with messaging platforms, executing local system commands, managing calendars, and orchestrating workflows across services. They are designed to operate persistently, retain long-term memory, and bridge natural language instruction with executable operations on behalf of human users. These capabilities represent a new class of agent behavior that transcends conventional prompt-response interactions and moves toward truly autonomous software assistants.

From the perspective of API discovery, such agents require reliable and comprehensive knowledge of the target system’s API surface in order to fulfill user intents accurately. Incomplete or outdated documentation—as exemplified by the shadow APIs in Dify—poses a significant obstacle for LLM-based agents that rely on documentation to generate correct invocation sequences and compose complex behaviors. If key interfaces are undocumented, agents may either be unable to perform certain tasks or may generate incorrect API invocations, leading to failed executions or unintended side effects. Furthermore, dynamic capabilities such as workflow orchestration, permission management, and contextual decision making demand fine-grained API discovery that goes beyond static endpoint descriptions.

The Clawbot example illustrates a broader trend in which autonomous agents are increasingly expected to handle real-world tasks and integrate with existing digital ecosystems in a seamless manner. Agents that can access multiple platforms (e.g., messaging apps, local system resources, external services) and execute multi-step operations inherently depend on accurate API knowledge to translate high-level user goals into low-level actions. Therefore, methodologies like traffic-based API discovery, which can systematically uncover both documented and shadow APIs, provide a critical foundation for empowering LLM-based agents to operate reliably in complex environments.

Looking forward, comprehensive API discovery and modeling will enable several key advancements for agent-centric systems. First, it will improve the automation capabilities of agents, allowing them to perform tasks such as end-to-end workflow execution, proactive monitoring, and service configuration without manual intervention. Second, it will facilitate safer and more predictable agent behavior by reducing reliance on incomplete or inaccurate documentation. Third, accurate API discovery will support the development of higher-order reasoning and planning capabilities in agents, enabling them to sequence dependent operations, handle error conditions, and adapt to evolving application surfaces. Ultimately, integrating robust API discovery into agent frameworks will be essential for realizing the promise of autonomous software agents that operate as reliable collaborators across diverse software ecosystems.

\section{Limitations and Future Work}

Despite the promising results achieved by \APIGH, our study has several limitations that point to important directions for future research. The primary limitation lies in the end-to-end completeness of runtime traffic collection. While \APIGH is capable of accurately discovering APIs from observed traffic, the effectiveness of the discovery process is inherently bounded by the coverage of the collected runtime interactions.

In our current experimental setup, runtime traffic was generated through manual and semi-natural interactions, such as daily usage, exploratory clicking, and execution of common workflows within the target web applications. Although this approach reflects realistic usage scenarios and captures a diverse set of API interactions, it does not guarantee exhaustive coverage of all reachable API endpoints. As a result, APIs that are guarded by rare conditions, deep interaction paths, or specific user roles may remain undiscovered. Achieving fully end-to-end API discovery therefore requires not only robust traffic analysis, but also systematic and automated generation of triggering interactions that exercise the full functional surface of the application.

Automated interaction and test generation represents a promising complementary direction to address this limitation. Recent work such as APICARV~\cite{yandrapally2023apicarv} demonstrates the feasibility of mining executable interaction paths from existing test cases to systematically explore API behaviors. Integrating such automated test-driven or behavior-driven interaction generation techniques with traffic-based API discovery could significantly improve coverage and reduce reliance on manual exploration. In particular, combining automated click-path generation with runtime traffic analysis may enable scalable, end-to-end discovery of web APIs across diverse applications with minimal human intervention.

Looking forward, we envision a unified framework in which automated interaction generation continuously stimulates application behaviors, while traffic-based analysis incrementally reconstructs and refines the API surface. Such an approach would further enhance the generality and practicality of black-box API discovery, making it applicable to large-scale, continuously evolving web systems. Exploring the integration of automated testing, agent-based exploration, and traffic-based API inference remains an important avenue for future work.

\section{Threats to Validity}
\label{sec:TTV}

\subsection{Construct Validity}
A threat to construct validity arises from the construction of the runtime traffic dataset. The traffic traces are manually collected through controlled executions, which may not exhaustively cover all possible API endpoints or rare execution paths. To mitigate this threat, we repeatedly exercised core functionalities of each application and validated collected traffic against official documentation and observed runtime behavior. Only endpoints that were confirmed to be callable and semantically meaningful were annotated as valid APIs. Although the dataset may be incomplete, the collected traffic and labels are accurate and sufficient for evaluating API discovery precision and robustness, which are the primary goals of this study.

\subsection{Internal Validity}
Internal validity may be affected by design choices in preprocessing and clustering, such as traffic filtering, path normalization, and parameter settings, which could influence the discovery results. To reduce this risk, all preprocessing steps are applied consistently, and baseline methods are evaluated under their commonly used configurations without dataset-specific tuning. In addition, we conduct ablation studies to examine the impact of each major component in \APIGH, showing that the performance improvements cannot be attributed to a single heuristic or implementation artifact.

\subsection{External Validity}
The external validity of our results may be limited by the scope of evaluated applications and experimental environments. Although we study six real-world web applications from different domains, the findings may not generalize to all systems, such as highly customized enterprise applications or those using non-standard communication protocols. Moreover, real-world traffic collection environments may introduce more complex noise than our simulated settings. To address this concern, we evaluate robustness under multiple noise types and levels, and observe stable performance trends, suggesting that the conclusions are likely to hold under practical deployment conditions, albeit with potential minor performance variation.

\section{Conclusion}
\label{sec: conclusion}
In this paper, we propose \APIGH, a black-box API discovery framework that reconstructs application APIs from mixed runtime traffic under unsupervised settings. By integrating traffic denoising with a two-stage clustering design based on structural templates and graph-based refinement, \APIGH reduces false positives and improves discovery precision in heterogeneous traffic environments.
Experiments on six real-world web applications show that \APIGH achieves state-of-the-art accuracy while maintaining strong robustness under different noise conditions. Ablation studies further confirm the necessity of each design component. In addition, our evaluation reveals inconsistencies between official API documentation and actual API usage in one application, demonstrating the practical value of runtime-based API discovery.
Overall, \APIGH provides an effective solution for accurate API discovery in closed-source and mixed-traffic environments.

\section*{Data Availability}

All source code and datasets used in this study are open source and anonymous. The source code and datasets are released at: \url{https://figshare}

\bibliographystyle{ACM-Reference-Format}
\bibliography{APIsense}

@article{Murphy2017,
  title={Preliminary analysis of REST API style guidelines},
  author={Murphy, Lauren and Alliyu, Tosin and Macvean, Andrew and Kery, Mary Beth and Myers, Brad A},
  journal={Ann Arbor},
  volume={1001},
  year={2017},
  publisher={}
}

@article{dai2022logram,
  title = {Logram: Efficient Log Parsing Using n-Gram Dictionaries},
  author = {Dai, Haibo and Li, Heng and Chen, Chih-Sheng and Shang, Weiyi and Chen, Tse-Hsun},
  journal = {IEEE Transactions on Software Engineering},
  volume = {48},
  year = {2022},
  publisher={}
}

@article{wang2024dainfer,
  author = {Wang, Chengpeng and Zhang, Jipeng and Wu, Rongxin and Zhang, Charles},
  title = {{DAInfer}: Inferring {API} Aliasing Specifications from Library Documentation via Neurosymbolic Optimization},
  journal = {Proceedings of the ACM on Software Engineering (FSE)},
  volume = {1},
  year = {2024},
  publisher = {ACM}
}

@inproceedings{8530111,
  author={Peng, Xin and Zhao, Yifan and Liu, Mingwei and Zhang, Fengyi and Liu, Yang and Wang, Xin and Xing, Zhenchang},
  title={Automatic Generation of API Documentations for Open-Source Projects}, 
  booktitle={2018 IEEE Third International Workshop on Dynamic Software Documentation (DySDoc3)}, 
  pages={7--8},
  publisher={IEEE},
  year={2018}
}

@inproceedings{espinha2014web,
  author = {Espinha, Tiago and Zaidman, Andy and Gross, Hans-Gerhard},
  title = {Web {API} growing pains: Stories from client developers and their code},
  booktitle = {Proceedings of the 18th European Conference on Software Maintenance and Reengineering (CSMR-WCRE)},
  pages = {84--93},
  publisher={IEEE},
  year = {2014}
}

@inproceedings{wang2019attributed,
  author = {Wang, Chun and Pan, Shirui and Hu, Ruiqi and Long, Guodong and Jiang, Jing and Zhang, Chengqi},
  title = {Attributed Graph Clustering: A Deep Attentional Embedding Approach},
  booktitle = {Proceedings of the 28th International Joint Conference on Artificial Intelligence (IJCAI)},
  pages = {3670--3676},
  publisher={IJCAI},
  year = {2019}
}

@inproceedings{Nybom2018,
  title={A systematic mapping study on API documentation generation approaches},
  author={Nybom, Kristian and Ashraf, Adnan and Porres, Ivan},
  booktitle={2018 44th euromicro conference on software engineering and advanced applications (SEAA)},
  pages={462--469},
  publisher={IEEE},
  year={2018}
}

@inproceedings{learning_glossary,
  title={A learning-based approach for automatic construction of domain glossary from source code and documentation},
  author={Wang, Chong and Peng, Xin and Liu, Mingwei and Xing, Zhenchang and Bai, Xuefang and Xie, Bing and Wang, Tuo},
  booktitle={Proceedings of the 2019 27th ACM joint meeting on european software engineering conference and symposium on the foundations of software engineering},
  pages={97--108},
  publisher={ACM},
  year={2019}
}

@inproceedings{Huang2018,
  title={API method recommendation without worrying about the task-API knowledge gap},
  author={Huang, Qiao and Xia, Xin and Xing, Zhenchang and Lo, David and Wang, Xinyu},
  booktitle={Proceedings of the 33rd ACM/IEEE International Conference on Automated Software Engineering},
  pages={293--304},
  publisher={ACM},
  year={2018}
}

@inproceedings{Mcmillan2011,
  title={Portfolio: finding relevant functions and their usage},
  author={McMillan, Collin and Grechanik, Mark and Poshyvanyk, Denys and Xie, Qing and Fu, Chen},
  booktitle={Proceedings of the 33rd International Conference on Software Engineering},
  pages={111--120},
  publisher={ACM},
  year={2011}
}

@inproceedings{Chen2014,
  title={Who asked what: Integrating crowdsourced faqs into api documentation},
  author={Chen, Cong and Zhang, Kang},
  booktitle={Companion Proceedings of the 36th International Conference on Software Engineering},
  pages={456--459},
  publisher={ACM},
  year={2014}
}

@inproceedings{Rahman2016,
  title={Rack: Automatic api recommendation using crowdsourced knowledge},
  author={Rahman, Mohammad Masudur and Roy, Chanchal K and Lo, David},
  booktitle={2016 IEEE 23rd International Conference on Software Analysis, Evolution, and Reengineering (SANER)},
  pages={349--359},
  publisher={IEEE},
  year={2016}
}

@inproceedings{wang2023gdoc,
  title={gDoc: Automatic generation of structured API documentation},
  author={Wang, Shujun and Tian, Yongqiang and He, Dengcheng},
  booktitle={Companion Proceedings of the ACM Web Conference 2023},
  pages={53--56},
  publisher={ACM},
  year={2023}
}

@inproceedings{nedelkoski2020log,
  title={Self-supervised log parsing},
  author={Nedelkoski, Sasho and Bogatinovski, Jasmin and Acker, Alexander and Cardoso, Jorge and Kao, Odej},
  booktitle={Joint European Conference on Machine Learning and Knowledge Discovery in Databases},
  pages={122--138},
  publisher={Springer},
  year={2020}
}

@inproceedings{Gu2016deep,
  title={Deep API learning},
  author={Gu, Xiaodong and Zhang, Hongyu and Zhang, Dongmei and Kim, Sunghun},
  booktitle={Proceedings of the 2016 24th ACM SIGSOFT international symposium on foundations of software engineering},
  pages={631--642},
  publisher={ACM},
  year={2016}
}

@inproceedings{liu2020webapisearch,
  title = {Web API Search: Discover Web API and Its Endpoint with Natural Language Queries},
  author = {Liu, Li and Bahrami, Mohammad and Park, Jongmin and Chen, Wen-Pin},
  booktitle = {Proceedings of the International Conference on Web Services (ICWS)},
  pages={},
  publisher={IEEE},
  year = {2020}
}

@inproceedings{liu2022uniparser,
  title = {UniParser: A Unified Log Parser for Heterogeneous Log Data},
  author = {Liu, Yiming and Zhang, Xu and He, Shilin and Zhang, Haixun and Li, Li and Kang, Yang and Xu, Yifan and Ma, Ming and Lin, Qingwei and Dang, Yingnong and Rajmohan, Saravanakumar and Zhang, Dongmei},
  booktitle = {Proceedings of The ACM Web Conference 2022},
  pages = {1893--1901},
  publisher={ACM},
  year = {2022}
}

@inproceedings{vaarandi2015logcluster,
  title = {LogCluster: A Data Clustering and Pattern Mining Algorithm for Event Logs},
  author = {Vaarandi, Risto and Pihelgas, Margus},
  booktitle = {Proceedings of the 11th International Conference on Network and Service Management (CNSM)},
  pages={},
  publisher={IEEE},
  year = {2015}
}

@inproceedings{yandrapally2023apicarv,
  title = {Carving UI Tests to Generate API Tests and API Specification},
  author = {Yandrapally, Raghavendra and Sinha, Shreya and Tzoref-Brill, Roni and Mesbah, Ali},
  booktitle = {Proceedings of the International Conference on Software Engineering (ICSE)},
  pages={},
  publisher={IEEE},
  year = {2023}
}

@inproceedings{yang2018extracting,
  title = {Towards Extracting Web API Specifications from Documentation},
  author = {Yang, Jiamou and Wittern, Erik and Ying, Andrew T. T. and Dolby, Julian and Tan, Lin},
  booktitle = {Proceedings of the 15th International Conference on Mining Software Repositories},
  pages = {454--464},
  publisher={ACM},
  year = {2018}
}

@inproceedings{logppt,
  title={Log parsing with prompt-based few-shot learning},
  author={Le, Van-Hoang and Zhang, Hongyu},
  booktitle={2023 IEEE/ACM 45th International Conference on Software Engineering (ICSE)},
  pages={2438--2449},
  publisher={IEEE},
  year={2023}
}

@inproceedings{huang2024generating,
  author = {Huang, Ruidong and Motwani, Mohit and Martinez, Isabella and Orso, Anderson},
  title = {Generating {REST} {API} Specifications through Static Analysis},
  booktitle = {Proceedings of the 46th IEEE/ACM International Conference on Software Engineering},
  pages = {1--13},
  publisher = {ACM},
  year = {2024}
}

@inproceedings{shi2011empirical,
  title={An empirical study on evolution of API documentation},
  author={Shi, Lin and Zhong, Hao and Xie, Tao and Li, Mingshu},
  booktitle={International Conference on Fundamental Approaches To Software Engineering},
  pages={416--431},
  year={2011},
  organization={Springer}
}

@inproceedings{wang2025codesync,
  title={CodeSync: Synchronizing Large Language Models with Dynamic Code Evolution at Scale},
  author={Wang, Chenlong and Chu, Zhaoyang and Cheng, Zhengxiang and Yang, Xuyi and Qiu, Kaiyue and Wan, Yao and Zhao, Zhou and Shi, Xuanhua and Jin, Hai and Chen, Dongping},
  booktitle={International Conference on Machine Learning},
  pages={62672--62700},
  year={2025},
  organization={PMLR}
}

@article{basu2026openclaw,
  title={OPENCLAW CHATBOTS ARE RUNNING AMOK: SCHOLARS ARE LISTENING},
  author={Basu, Mohana},
  journal={Nature},
  volume={650},
  pages={533},
  year={2026}
}

@inproceedings{wolflein2025agenttools,
  title={LLM Agents Making Agent Tools},
  author={Wölflein, Georg and others},
  booktitle={Proceedings of ACL},
  year={2025}
}

@inproceedings{zhong2013detecting,
  title={Detecting API documentation errors},
  author={Zhong, Hao and Su, Zhendong},
  booktitle={Proceedings of the 2013 ACM SIGPLAN international conference on Object oriented programming systems languages \& applications},
  pages={803--816},
  year={2013}
}

@inproceedings{luo2024holistic,
  author = {Luo, Chengpeng and Ouyang, Yicheng and Zhao, Yanyang and Zhang, Jian and Jiang, Yu and Yang, Zijiang},
  title = {Holistic Concolic Execution for Dynamic Web Applications via Symbolic Interpreter Analysis},
  booktitle = {Proceedings of the 2024 ACM SIGSAC Conference on Computer and Communications Security},
  pages={},
  publisher = {ACM},
  year = {2024}
}

@inproceedings{lin2023detecting,
  author = {Lin, Zhiqiang and Liu, Xiangyuan and Wang, Jianing and Liang, Zhenkai},
  title = {Detecting {API} Post-Handling Bugs Using Code and Inference},
  booktitle = {Proceedings of the 32nd USENIX Security Symposium},
  pages = {1--18},
  publisher = {USENIX Association},
  year = {2023}
}

@inproceedings{wallner2024feature,
  author = {Wallner, Felix and Aichernig, Bernhard A. and Burghard, Christian},
  title = {It's Not a Feature, It's a Bug: Fault-Tolerant Model Mining from Noisy Data},
  booktitle = {Proceedings of the 46th IEEE/ACM International Conference on Software Engineering},
  pages = {1--12},
  publisher = {ACM},
  year = {2024}
}

@inproceedings{wang2024diagnosing,
  author = {Wang, Zhe and Hu, Huanwu and Kong, Linghe and others},
  title = {Diagnosing Application-network Anomalies for Millions of {IPs} in Production Clouds},
  booktitle = {Proceedings of the 2024 USENIX Annual Technical Conference},
  pages={},
  publisher = {USENIX Association},
  year = {2024}
}

@inproceedings{han2024byways,
  author = {Han, Xinyu and Gao, Yuan and Parmer, Gabriel and Wood, Timothy},
  title = {Byways: High-Performance, Isolated Network Functions for Multi-Tenant Cloud Servers},
  booktitle = {Proceedings of the 2024 ACM Symposium on Cloud Computing},
  pages={},
  publisher={ACM},
  year = {2024}
}

@inproceedings{vanede2020flowprint,
  author = {Van Ede, Tim and Bortolameotti, R. and others},
  title = {{FlowPrint}: Semi-Supervised Mobile-App Fingerprinting on Encrypted Network Traffic},
  booktitle = {Proceedings of the 27th Annual Network and Distributed System Security Symposium},
  pages={},
  publisher={},
  year = {2020}
}

@misc{openapi_spec,
  author = {{OpenAPI Initiative}},
  title = {{OpenAPI Specification}},
  year = {2023},
  url = {https://www.openapis.org}
}

@misc{drain3_github,
  author = {Logpai},
  title = {{Drain3}: A robust streaming log template miner},
  year = {2019},
  url = {https://github.com/logpai/Drain3}
}

@misc{optic,
  title = {Optic: API Specification Generation and Validation from Traffic},
  howpublished = {\url{https://github.com/opticdev/optic}},
  note = {Open-source tool}
}

@misc{mitmproxy2swagger,
  title = {mitmproxy2swagger: Automatically Convert Mitmproxy Captures to OpenAPI},
  howpublished = {\url{https://github.com/alufers/mitmproxy2swagger}},
  note = {Open-source tool}
}

\end{document}